\documentclass[a4paper,11pt]{article}
\pdfoutput=1 

\usepackage{jcappub} 

\usepackage[T1]{fontenc} 

\usepackage{multirow}
\usepackage{adjustbox}
\usepackage{hyperref}
\title{\boldmath Constraining the hadronic properties of star-forming galaxies above $1\, \rm GeV$ with 15-years Fermi-LAT data}

\author[a,b,1]{A. Ambrosone,\note{Corresponding author.}}
\author[c,d]{M. Chianese}
\author[c,d,e]{A. Marinelli}

\affiliation[a]{Gran Sasso Science Institute (GSSI), Viale Francesco Crispi 7, 67100 L’Aquila, Italy}
\affiliation[b]{ INFN-Laboratori Nazionali del Gran Sasso(LNGS), via G. Acitelli 22, 67100 Assergi (AQ), Italy}
\affiliation[c]{Dipartimento di Fisica ``Ettore Pancini'', Università degli studi di Napoli ``Federico II'', Complesso Univ. Monte S. Angelo, I-80126 Napoli, Italy}
\affiliation[d]{INFN - Sezione di Napoli, Complesso Univ. Monte S. Angelo, I-80126 Napoli, Italy}
\affiliation[e]{INAF - Osservatorio Astronomico di Capodimonte, Salita Moiariello 16, I-80131 Naples, Italy}
\emailAdd{antonio.ambrosone@gssi.it}
\emailAdd{marco.chianese@unina.it}
\emailAdd{antonio.marinelli@unina.it}

\abstract{ Star-forming and starburst galaxies (SFGs and SBGs) are considered to be powerful emitters of non-thermal $\gamma$-rays and neutrinos, due to their intense phases of star-formation activity, which should confine high-energy Cosmic-Rays (CRs) inside their environments. On this regard, the Fermi-LAT collaboration has found a correlation between the $\gamma$-ray and infrared luminosities for a sample of local sources. Yet, the physics behind these non-thermal emission is still under debate. We provide novel constraints on the tight relation between $\gamma$-rays and star formation rate (SFR) exploiting 15 years of public Fermi-LAT data. Thus, we probe the calorimetric fraction~$F_{\rm cal}$ of high-energy protons in SFGs and SBGs, namely, the fraction of high-energy protons actually producing high-energy $\gamma$-rays and neutrinos. Further, we extrapolate this information to their diffuse $\gamma$-ray and neutrino emissions constraining their contribution to the extra-galactic gamma-ray background (EGB) and the diffuse neutrino flux. Using the publicly-available \texttt{fermitools},  we analyse 15.3 years of $\gamma$-ray between $1-1000\, \rm GeV$ data for 70 sources, 56 of which were not previously detected. We relate this emission to a theoretical model for SBGs in order to constrain $F_{\rm cal}$ for each source and then study its correlation with the star formation rate of the sources. Firstly, we find at~$4\sigma$ level an indication of $\gamma$-ray emission for other two SBGs, namely M 83 and NGC 1365. By contrast, we find that, even with the new description of background, the significance for the $\gamma$-ray emission of M 33~(initially reported as discovered) still stands at $\sim \, 4\sigma$ (as already reported by previous works). 
Along with previous findings, the flux of each detected source is consistent with a $\sim E^{-2.3/2.4}$ spectrum, compatible with the injected CR flux inferred for CRs in the Milky-Way. We  also notice that the correlation between $F_{\rm cal}$ and the SFR is in accordance with the expected scaling relation for CR escape dominated by advection. We remark that undiscovered sources strongly constrain $F_{\rm cal}$ at 95\% CL, providing fundamental information when we interpret the results as common properties of SFGs and SBGs. Finally, we find that these sources might contribute $(12\pm 3)\%$ to the EGB, while the corresponding diffuse neutrino flux strongly depends on the spectral index distribution along the source class.}

\begin{document}
\maketitle
\flushbottom

\section{Introduction}

Star-forming and starburst galaxies (SFGs and SBGs) are galaxies in a phase of intense star formation, leading to high gas density and to an enhanced rate of supernovae (SN) explosions~\cite{Peretti:2018tmo}. This activity is expected to be directly correlated to $\gamma$-rays and neutrinos, via proton-proton (pp) collisions between high-energy Cosmic-Rays (CRs) accelerated by supernovae remnants (SNRs) and the gas~\cite{Peretti:2018tmo,Peretti:2019vsj,Ambrosone:2021aaw,Ambrosone:2020evo,2020A&A...641A.147K,Kornecki:2021xiy,Merckx:2023kvn,Werhahn:2021bal,Werhahn:2021jvy,Werhahn:2023osl,Nunez-Castineyra:2022rjc}.
The Fermi-LAT collaboration has indeed detected a sample of 14 SFGs which feature a correlation between the $\gamma$-ray luminosity from $100\, \rm MeV$ to $100\,\rm GeV$ $(L_{[0.1-100]\, \rm GeV})$ and the infrared luminosity $(L_{8-1000\, \mu \rm m})$ \cite{2012ApJ...755..164A}. These detections have also been updated by several authors, such as \citep{Rojas-Bravo:2016val,Ajello:2020zna,Xiang:2023dww}.

These results are typically interpreted as an evidence for the existence of common properties shared by the entire population of SFGs and SBGs~\citep{Kornecki:2021xiy}.
For instance, the fact that these sources present hard power-law spectra $E^{-\gamma}$, with $\gamma \sim 2.2-2.3$, might indicate that the physics of CRs is dominated by energy-independent mechanisms, such as the pp inelastic timescale or advection~\cite{Lacki:2013ry}.
However, some works~\cite{Krumholz:2019uom,Roth:2021lvk,Roth:2022hxc} have recently proposed that CR transport inside these source might be dominated by diffusion for $E_{\rm CR} \ge 100\, \rm GeV$, leading to a suppression of the $\gamma$-ray and neutrino production rates at higher energies.
Physically, this means that the calorimetric fraction~$F_{\rm cal}$ -- i.e. the fraction of high-energy CRs which actually lose energy inside SFGs and SBGs producing $\gamma$-rays and neutrinos -- might be smaller than previously predicted and also energy-dependent. 
In order to discriminate these two scenarios, however, new and more precise measurements (especially in the TeV energy range) are required~\cite{Ambrosone:2022fip}.
Several authors have attempted in modelling the calorimetric fraction along a large star formation rate (SFR) range~$(10^{-2}-10^{3}\, \rm M_{\odot}\, \rm yr^{-1})$
~\cite{2012ApJ...755..164A,2020A&A...641A.147K,Kornecki:2021xiy,Roth:2022hxc,Werhahn:2021bal,Werhahn:2021jvy,Werhahn:2023osl,Nunez-Castineyra:2022rjc,Pfrommer:2017jau,Crocker:2020yub}. All the theoretical studies point towards the conclusion that $F_{\rm cal}$ increases with SFR and so CRs lose most of their energy in the environment of SBGs. However, the actual degree of calorimetry is still under debate due to a lack of knowledge of the CR escape mechanisms from these astrophysical environments~\cite{Kornecki:2021xiy}. 
In this paper, we provide new data-driven constraints on the calorimetric fraction of SFGs and SBGs by analysing a catalogue of 70 sources introduced by \cite{2012ApJ...755..164A}, using $\sim 15.3\, \rm years$ of Fermi-LAT data\footnote{Fermi-LAT data can be freely downloaded at \url{https://fermi.gsfc.nasa.gov/ssc/data/access/}} and the publicly-available \texttt{fermitools}.\footnote{The \texttt{fermitools} are available at~\url{https://fermi.gsfc.nasa.gov/ssc/data/analysis/software/}} 
In particular, we search for $\gamma$-ray emission between $1\, \rm GeV$ and $1\, \rm TeV$, dividing the catalogue in two samples: the 56 sources not yet detected~\footnote{In this paper, we denote discovered sources as those exceeding the Fermi-LAT discovery threshold.} and the 14 sources which have been previously detected. 
We find strong hints of $\gamma$-ray emission in coincidence of M 83 and NGC 1365 at level of $\sim 4\sigma$. Furthermore, for M 33 which was previously reported as a discovered source by~\cite{Ajello:2020zna}, we find that its gamma-ray emission still stands right below the $4\sigma$ detection level (as already emphasised by \cite{Nunez-Castineyra:2022rjc,Werhahn:2023osl}).


Then, we test a physically-motivated relation between $F_{\rm cal}$ and the rate of supernovae explosion $R_{\rm SN}$, in contrast with the simplistic power-law function previously tested~\cite{2012ApJ...755..164A,Rojas-Bravo:2016val,Ajello:2020zna,Xiang:2023dww}, finding a good agreement with the data. We emphasise that the correct estimate of the systematic uncertainty on $\rm R_{\rm SN}$ is crucial in order to extract the correct information on this correlation. Moreover, undiscovered sources place strong constraints to $F_{\rm cal}$, thus slightly modifying the $F_{\rm cal}$--$R_{\rm SN}$ correlation. Therefore, in order to interpret these emissions as shared properties of all SFGs and SBGs is also important to take into account sources which present no evidence for $\gamma$-ray emission. 

Finally, we employ such a correlation to evaluate the diffuse $\gamma$-rays and neutrinos flux from the whole source population. In order to do this, we make use of the recently-updated cosmic star-formation-rate distribution obtained through the James Webb Space Telescope reported by~\cite{Kim:2023xxh}. We find that SFGs and SBGs might contribute $(12\pm 3)\%$ to the extragalactic gamma-ray background (EGB)~\cite{Fermi-LAT:2014ryh} above $50\, \rm GeV$, while their contribution to the diffuse neutrino flux measured by IceCube with 6-year cascade events~\cite{IceCube:2020acn} might vary from 4\% to 18\% crucially depending on the assumed distribution of the spectral indexes along the source class. 

The paper is structured as follows. In Sec.~\ref{sec:sample_galaxies}, we describe the sample of galaxies analysed. In Secs.~\ref{sec:data_analysis} and~\ref{sec:results_statistical_analysis}, we describe the statistical analysis of the Fermi-LAT data and report the corresponding results, respectively. In Sec.~\ref{sec:theoretical_model}, we discuss the theoretical model we adopt to evaluate the $\gamma$-ray and neutrino fluxes from each source. In Sec.~\ref{sec:correlation_Fcal_Rsn}, we describe the $F_{\rm cal}$--$R_{\rm SN}$ correlation and discuss our findings. In Sec.~\ref{sec:diffuse}, we extrapolate our results to the diffuse $\gamma$-ray and neutrino fluxes. Finally, in Sec.~\ref{sec:conclusions}, we draw our conclusions. The paper has five appendices: in appendix~\ref{app:SED}, we report all the new spectral energy distributions (SEDs) for the sources above the $5\sigma$ discovery threshold; in appendix~\ref{app:diffuse_flux}, we discuss the properties of the diffuse spectrum; in appendix~\ref{app:Rsn_systematic_uncertainty} we comment on the impact of the systematic uncertainty affecting $R_{\rm SN}$ on our results; in appendix~\ref{app:power-law}, we discuss the  fit of data   with  a power-law function and the comparison with the function used in the main text and finally in appendix~\ref{app:AGN_contamination}, we discuss on the impact of the sources with potential AGN contamination in the data fit presented in the main text.

\section{Sample of galaxies}\label{sec:sample_galaxies}

We investigate the gamma-ray emission of 70 sources which we divide into two samples:
\begin{itemize}
    \item \textbf{Sample A} (see Tab.~\ref{tab:sample_A}): it contains the galaxies introduced by~\cite{2012ApJ...755..164A} (see also~\cite{Rojas-Bravo:2016val}) for which  no $\gamma$-ray detection has been reported yet. These galaxies exhibit a galactic latitude coordinate $|\rm b| \ge 10^{\circ}$ and, therefore, the contamination from the diffuse galactic $\gamma$-ray emission is negligible. For these sources, we take the distances and the total infrared luminosity between $8-1000\, \rm \mu \rm m$ $(L_{\rm IR})$ from~\cite{Rojas-Bravo:2016val}, consistently rescaled for the different hubble parameter~$\rm H_0$ used.\footnote{In this work, we adopt the value $67.74\, \rm km\, \rm s^{-1}\, \rm Mpc^{-1}$.}
    \item \textbf{Sample B} (see Tab.~\ref{tab:sample_B}): it refers to the 14 sources discovered in $\gamma$-rays reported by \cite{Ajello:2020zna}, including also the Circinus Galaxy reported by \cite{2020A&A...641A.147K,Kornecki:2021xiy}. For this sample, we use the distances and the infrared luminosity reported by \cite{2020A&A...641A.147K}. For NGC 3424, ARP 220 and ARP 299, we use the updated values reported by \cite{Nunez-Castineyra:2022rjc}.
\end{itemize}
Some of these sources are not only classified as SFGs but also AGNs, with Seyfert activity. For this reason, we focus on $\gamma$-ray emission above $E \ge 1\, \rm GeV$, where the photons from seyfert activity are expected to be negligible~\citep{Inoue:2019fil,Murase:2023ccp,Kheirandish:2021wkm}.

\section{Data analysis}\label{sec:data_analysis}

We analyse the latest Fermi-LAT data which have been collected in sky-survey mode from August 2008 and November 2023, from a Mission Elapsed Time  239557417 s to 720724699 s, with a total lifetime of $\sim$$15.3 \, \rm years$. We select photons in the energy range $[1-1000]\, \rm GeV$, which strongly reduces the possibility of mis-identification of sources due to a limited PSF dimension of $\sim$$10^{\circ}$ at lower energies.  We consider events belonging to  \texttt{P8R3\_v3} version of the Pass 8 photon dataset and the corresponding \texttt{P8R3\_SOURCE\_V3} instrument response functions. Data are analysed using the publicly-available \texttt{fermitools} provided by the Fermi-LAT collaboration and their analysis threads.\footnote{The \texttt{fermitools} analysis threads are available at \url{https://fermi.gsfc.nasa.gov/ssc/data/analysis/scitools/}}
We consider a $15^{\circ} \times 15^{\circ}$ Region of Interest (RoI) centred at the equatorial coordinates of each source, selecting only the data passing the filter for being considered of good-quality \texttt{(DATA\_QUAL>0)\&\&(LAT\_CONFIG==1)}. 

In order to reduce the contamination from the Earth's limb, following the default suggestions in the \texttt{fermitools}, the events with zenith angle $>$$90^{\circ}$ are excluded. We emphasise that the Fermi-LAT collaboration has recently updated the selection for events above $1\, \rm GeV$, selecting events for zenith angle $<$$105^{\circ}$~\citep{Fermi-LAT:2019yla}. However, we have verified that the results do not change either for sample A or for sample B, even with this new selection. Therefore, we prefer to leave the event selection suggested in the \texttt{fermitools} in order to work with a photon sample with higher purity.

These data are analysed following the binned maximum likelihood ratio method, which is officially released by the Fermi-LAT collaboration. The likelihood function is defined as~\citep{Malyshev:2023xya}
\begin{equation}
    \mathcal{L} = \prod_{i} \mathcal{P}\left(  E_i ,\, X_i \,|\, M_{i}(\Omega)\right)
\end{equation}
where $ \mathcal{P}( E_i ,\, X_i \,|\, M_{i}(\Omega))$ is the Poisson probability distribution function for observing a photon of a given energy $E_i$ and direction $X_i$, given the expected number of photons $M_i$ provided by the model which depends on the $\Omega$ parameters. The index $i$ runs over the bins for the events in the RoI. We determine the test statistic $\rm TS$ for each source as 
\begin{equation}\label{eq:lik}
    {\rm TS} = -2\,\ln \frac{\mathcal{L}_0}{\mathcal{L}}
\end{equation}
where $\mathcal{L}_0$ is the maximised likelihood in the background-only hypothesis, namely in the hypothesis the source does not emit photons, and $\mathcal{L}$ is the maximised likelihood including the source under study. 
The conversion from the TS to the significance level can be performed using a chi-squared $\chi^2$ distribution with degrees of freedom equal to the number of the free parameters for the source model~\citep{Malyshev:2023xya}. For instance, for power-law spectra, considering both normalisation and spectral index as free parameters, $\rm TS = 25$ (also defined as discovery threshold for the TS) corresponds to $\sim 4.6 \sigma$ significance.  

In order to maximise the likelihood in Eq.~\eqref{eq:lik}, the data count maps are binned in angular coordinates, with $0.1^{\circ}$ bin per pixel, and in energy with 37 logarithmically spaced bins.~\footnote{The Analysis threads of the \texttt{fermitools} advise of using at least 10 bins per decade. Since we analyse exactly 3 decades, we leave the default value of 37 energy bins.}
The background hypothesis comprises all the sources in the 4FGL catalogue \texttt{gll\_psc\_v32.xml}~\cite{Fermi-LAT:2022byn,Ballet:2023qzs}, the standard isotropic extragalactic emission \texttt{iso\_P8R3\_SOURCE\_V3\_v1} and the galactic diffuse emission \texttt{gll\_iem\_v07}. In order to account for the finite dimension of the PSF, we also consider sources outside the RoI with a further radius of $5^{\circ}$. As suggested by~\cite{Malyshev:2023xya}, the fit is performed in an iterative way and at each step sources with very low $\rm TS$, such as spurious solutions with $\rm TS < 0$, are eliminated from the likelihood. 

In this work, for the signal hypothesis, we test power-law spectra $\phi_\gamma = \phi_0 \,E^{-\gamma}$ added at the nominal position of the source\footnote{By nominal position, we mean that for sources of sample A, we use the NED position of the sources, while for sources of sample B, we use the default position available in the 4FGL catalogue}. In the likelihood maximisation, we fit all the sources leaving free the source parameters (normalisation $\phi_0$ and spectral index $\gamma$) within $5^{\circ}$ of the RoI centre. Furthermore, we leave free the normalisation of extremely variable sources up to $15^{\circ}$ of the RoI centre~\footnote{please see \url{https://github.com/physicsranger/make4FGLxml}} as well as the normalisation for the isotropic extragalactic and the galactic diffuse templates. The other parameters are fixed to their best-fit values of the 4FGL catalogue. Finally, we also account for the energy dispersion using \texttt{edisp\_bins = -2} as advised in the \texttt{fermitools} threads. All the sources except for the Small Magellanic Cloud (SMC) and the Large Magellanic Cloud (LMC) are considered as point-like sources. For SMC and LMC, we instead utilise the official templates provided in the 4FGL catalogue. For these sources, we leave the source parameters to be free within $8^{\circ}$ and $6^{\circ}$ from the RoI centres, respectively.

\section{Results of the statistical analysis}\label{sec:results_statistical_analysis}

We report the obtained results in Tabs.~\ref{tab:sample_A} and~\ref{tab:sample_B} for the sample A and B, respectively.
For each source of the sample A for which the TS is smaller than the discovery threshold ($\rm TS < 25$), we report the luminosity distance, the infrared luminosity, and the $95\%$ CL upper limit on the flux in the range $1-1000\, \rm GeV$ assuming a spectral index $\gamma = 2.3$ as typical value for known SBGs (see results for the sample B). We do not find any excess, except for M83 and NGC 1365 which shows $\rm TS \sim 15$. For these cases, we also report the best-fit values and the $68.3\%$ CL limits in brackets. Differently from~\cite{Blanco:2023dfp}, we do not find any hint for NGC 3079: this is probably due to the fact that they look for photons with $E \ge 50\, \rm MeV$ where the limited Fermi-LAT PSF might cause mis-identification of sources. This problem has already been studied by~\cite{Rojas-Bravo:2016val} who pointed out that increasing the energy threshold leads to a better probe of the emission from single sources (and potentially reducing previous evidence of emission). Moreover, we find no evidence for $\gamma$-ray emission from the sources NGC 6946 and IC 342 which correlates with the most energetic CRs observed~\cite{TelescopeArray:2023sbd}.
\begin{table}[h!]
   \centering
   \begin{adjustbox}{max width=\textwidth, max height=0.45\textheight}
    \begin{tabular}{c|c|c|c|c}
      \multirow{2}{*}{Source}   & $D_L$ & $L_{\rm IR}$ & $F^{95\% \rm CL}_{1-1000\, \rm GeV} $ & \multirow{2}{*}{$\gamma$}\\ 
      & $\rm [Mpc]$ & $\rm [10^{10} \, L_{\odot}]$ & $\rm [10^{-11}\, \rm ph \, \rm cm^{-2}\, \rm s^{-1}]$ & \\ \hline
      NGC 3079   & 17.9 & 5.3 & 10.4 & 2.3 \\
      NGC 4631   &  8.97 & 2.45 & 5.3 & 2.3 \\
      M 83       & 4.1 & 1.72 & $17.7~(11 \pm 6)$ & $2.3~(2.2 \pm 0.2)$\\
      M 51 & 10.6 & 5.15 & 7.28 & 2.3 \\
      NGC 3628 & 8.4 & 1.22 & 5.81 & 2.3 \\
      NGC 4826 & 5.2 & 0.32 & 6.62 & 2.3 \\
      NGC 6946 & 6.1 & 1.96 & 2.16 & 2.3 \\
      NGC 2903 & 6.9 & 1.02 & 9.72 & 2.3 \\
      NGC 5055 & 8.1 & 1.35 & 7.91 & 2.3 \\
      IC 342 & 4.1 & 1.72 & 6.80 & 2.3 \\
      NGC 4414 & 10.3 & 0.99 & 3.96 & 2.3 \\
      NGC 891 & 11.4 & 3.19 & 13.8 & 2.3 \\
      NGC 3893 & 15.4 & 1.47 & 3.21 & 2.3 \\
      NGC 3556 & 11.7 & 1.72 & 3.67 & 2.3 \\
      NGC 1365 & 23.0 & $15.9$ & $12.4~(8\pm 3)$ & $2.3~(2.4 \pm 0.2)$ \\
      NGC 660 & 15.5 & 4.5 & 13.6 & 2.3 \\
      NGC 5005 & 15.5 & 1.71 & 6.26 & 2.3 \\
      NGC 1055 & 16.4 & 2.57 & 13.2 & 2.3 \\
      NGC 7331 & 16.6 & 4.29 & 10.3 & 2.3 \\
      NGC 4030 & 18.9 & 2.57 & 4.38 & 2.3 \\
      NGC 4041 &   19.9   &   2.08   & 5.50 & 2.3 \\
      NGC 1022 &   23.4       &   3.19       & 3.55 & 2.3 \\
      NGC 5775 &  23.6  &  4.66   &   5.01 & 2.3 \\
      NGC 5713 &  26.6 &5.15  &  1.55 & 2.3 \\
      NGC 5678 &  30.8  & 3.68    &  4.35 & 2.3 \\
      NGC 520  &   34.4    &  10.4     &   2.89 & 2.3 \\
      NGC 7479 &   39.0   &  9.1   &   11.2 & 2.3 \\
      NGC 1530 &   39.2     &  5.76 &    4.69 & 2.3 \\
      NGC 2276 &  39.3  &  7.60  &  5.26 & 2.3 \\
      NGC 3147 &  43.7 & 7.60   &   2.20 & 2.3 \\
      IC 5179 & 51.2   &  17.16  &  11.4 & 2.3 \\
      NGC 5135 &  57.2   &  17.16   &  8.89 & 2.3 \\
      NGC 6701 &  62.9 & 13.48   &   4.21 & 2.3 \\
      NGC 7771 &  66.9  &  25.7 & 8.18 & 2.3 \\  
      NGC 1614 & 70.0   &  47.8  &   10.5 & 2.3 \\
      NGC 7130 &   72.0  &  25.7  &   1.82 & 2.3 \\
      NGC 7469 & 74.7    &  50.3  &   3.94 & 2.3 \\
     IRAS 18293\, 3413 & 79.8   &  66.2 &  2.14 & 2.3 \\
     MRK 331 & 83.4  & 33.1 &   1.02 & 2.3 \\
      NGC 828 & 83.5 &  27.0    &  4.33 & 2.3 \\
      IC 1623 &   90.5  & 57.6  &  3.24 & 2.3 \\
      ARP 193 &  102.6  & 45.4  &  3.16 & 2.3 \\
      NGC 6240 &  108.6  &  74.8  &   3.68 & 2.3 \\
    NGC 1144 &   129.9  &  30.6   &  3.01 & 2.3 \\
    MRK 1027 &  136.7   &  31.9  &   2.84 & 2.3 \\
    NGC 695  &  147.8  & 57.6  &  6.58 & 2.3 \\
    ARP 148 &   158.7    &  44.1   &   8.18 & 2.3 \\
    MRK 273 &  168.5    &  159.4    &   4.77 & 2.3 \\
    UGC 05101 & 177.4   & 109.1  &  4.97 & 2.3 \\
    ARP 55 &  180.1   & 56.4   &  4.90 & 2.3 \\
    MRK 231 &   188.6  & 367.8   &   5.70 & 2.3 \\ 
    IRAS 05189\, 2524 &   188.6    &  147.1  &  5.08 & 2.3 \\
    IRAS 17208\, 0014 &   191.7   &  281.9 &   12.3 & 2.3 \\
    IRAS 10566+2448 & 191.9    & 115.3  &  3.76 & 2.3 \\
    VII Zw 31 &  247.3   & 106.6  &  1.69 & 2.3 \\
    IRAS 23365+3604 &  294.6  & 171.6   &   11.4 & 2.3 
       \end{tabular}
       \end{adjustbox}
    \caption{\label{tab:sample_A}Results for the sample A. From left to right: the source name, the luminosity distance, the infrared luminosity, the upper limit on the integrated $\gamma$-ray flux, the spectral index assumed to evaluate the upper limit. For hints of $\gamma$-ray emissions, we report the best-fit values in brackets.}
\end{table}


\begin{table}[h]
\centering
\begin{adjustbox}{max width=\textwidth}
    \begin{tabular}{c|c|c|c|c|c|c|c|}
    \multirow{2}{*}{Source}   & $D_{L}$ & $L_{\rm IR}$ & $F_{1-1000\, \rm GeV} $ & $\phi_0$ & \multirow{2}{*}{$\gamma$} & \multirow{2}{*}{$\rm TS~(\sigma)$} & \multirow{2}{*}{$\rm TS_{\rm SM}$} \\
    & $\rm [Mpc]$ & $\rm [10^{10} \, L_{\odot}]$ & $\rm [10^{-10}\, \rm ph \, \rm cm^{-2}\, \rm s^{-1}]$ & $[10^{-12}\, \rm MeV^{-1}\, \rm cm^{-2}\, \rm s^{-1}]$ & & & \\ \hline 
       M 82  & 3.53 & 5.6 &  $9.8 \pm 0.5$& $1.31 \pm 0.10$   & $2.34 \pm 0.06$ & 1104~(33)& 0.35\\
       NGC 253 & 3.56 &3.6 & $8.1 \pm 0.9$ & $1.08 \pm 0.10$ & $ 2.33 \pm 0.08$  & 730~(27)& 1.03 \\
       ARP 220 & 84.3 &$1.7\cdot 10^2$ & $1.6 \pm 0.6$ & $(2.0 \pm 0.7)\cdot 10^{-1}$ & $2.2 \pm 0.2$ & 50~(7.1)& -- \\
       NGC 1068 & 10.1 & 10.0 & $4.5\pm 0.5$ & $(5.8\pm 0.9)\cdot 10^{-1} $ & $2.28 \pm 0.15$ & 238~(15)& -- \\
       Circinus & 4.21 & 1.7& $5.1\pm 1.3$ & $(6.2 \pm 1.7)\cdot 10^{-1}$ & $2.23 \pm 0.14$ & 78~(8.8)& --\\
       SMC & 0.06 & $7.1\cdot 10^{-3}$ &  $(3.0 \pm 0.3)\cdot 10^{1}$ & $4.4 \pm 0.3$ & $2.44 \pm 0.06$ & 801~(28)& 4.13 \\
       M 31 & 0.77 & $2.3\cdot 10^{-1}$ & $3.1 \pm 0.8$ & $(6.3 \pm 1.3)\cdot 10^{-1}$ & $3.0 \pm 0.3$ & 74.6~(8.6)& 0.22\\
       NGC 2146 & 17.2 &12.6 & $1.3 \pm 0.5$ & $(1.5 \pm 0.5)\cdot 10^{-1}$ & $2.16 \pm 0.18$ & 41.5~(6.4)& -- \\
       ARP 299 & 48.6 & 72.6 & $1.3 \pm 0.5$ & $(1.7\cdot 0.6)\cdot 10^{-1}$ & $ 2.3 \pm 0.2$ & 46.4~(6.8)& -- \\
       NGC 4945 & 3.72 & 2.8 & $9.6 \pm 1.3$ & $1.34\pm 0.15$ & $2.40\pm 0.08$ & 412~(20)& -- \\
       NGC 2403 & 3.18 &0.15 & $1.5\pm 0.5$ & $(10\pm 4)\cdot 10^{-2}$ & $1.92 \pm 0.17$ & 52.8~(7.3)& -- \\
       NGC 3424 & 27.2 &2.1 & $10\pm 5$ & $(1.3 \pm 0.5)\cdot 10^{-1}$ & $2.3 \pm 0.3$ & 28~(5.3)& -- \\
      LMC & 0.05 &$5.2\cdot 10^{-2}$ & $(1.38 \pm 0.07)\cdot 10^{2}$ & $(1.85\pm 0.08)\cdot 10^{1}$ & $2.41\pm 0.04$ & 1493~(38)& 0.24 \\
      M 33 & 0.91 & 0.14& $1.2 \pm 0.6$$^\dagger$ & $(1.8\pm 0.7)\cdot 10^{-1}$ & $2.5\pm 0.3$ & 16~(4)& --
   
    \end{tabular}
       \end{adjustbox}
    \caption{Results for the sample B. From left to right: the source name, the luminosity distance, the infrared luminosity, the integrated $\gamma$-ray flux, the flux normalisation, the spectral index, the value of the test statistics. $^\dagger$Note that since M 33 is below the discovery threshold, we also compute the $95\%$ CL upper limit fixing $\gamma =2.3$, obtaining $F_{1-1000\, \rm GeV} = 1.65 \cdot 10^{-10}\, \rm ph \, \rm cm^{-2}\, \rm s^{-1}$.}
    \label{tab:sample_B}
\end{table}
For each source of the sample B, we also report the best-fit interval of the flux normalisation and spectral index at $68.3\%$ CL, and the corresponding value for the test statistics TS. Our results are in fair agreement with previous ones~\cite{Fermi-LAT:2017ztt,Ajello:2020zna,Fermi-LAT:2022byn}.
For SMC, we find a slightly softer spectrum than~\cite{Ajello:2020zna} being in agreement with~\cite{Fermi-LAT:2022byn}. 
For M31, along with the other sources of sample B, we have used the official point-like model present in the 4FGL catalogue, despite some other works have reported it as an extended source of $0.4^{\circ}$~\cite{Fermi-LAT:2017ztt,Ajello:2020zna}.
We obtain convergence anyway (with a $\rm TS \sim 75$), although with a very soft power-law spectrum $\sim E^{-3.0 \pm 0.3}$.
Finally, for M 33, there is not any match with sources present in the 4FGL catalogue. So, as for the sources in the sample A, we have added a point-like source in its position. Differently from~\cite{Ajello:2020zna}, we find only an excess with $({\rm TS} \sim 16)$ which is below the discovery threshold. In appendix~\ref{app:SED}, we report the SEDs for each source above the discovery threshold according to our analysis.
We stress that some of the sources of sample B (LMC, SMC, M31, M82, NGC253) are not reported as simple power-laws in the 4FGL catalogue. Therefore, we show that there is not a statistical difference in using those signal models as opposed to simple power-laws. To this purpose, we define 
\begin{equation}
    \rm TS_{\rm SM} = 2\cdot \rm Ln \bigg(\frac{\mathcal{L}_{\rm 4FGL}}{\mathcal{L}_{\rm PL}}\bigg)
\end{equation}
where $\mathcal{L}_{\rm 4FGL}$ is the likelihood maximised using the signal model in the 4FGL catalogue, while $\mathcal{L}_{\rm PL}$ is the likelihood maximised in the power-law model. We find that all the $\rm TS_{\rm SM}$ are much below the discovery threshold and so our signal assumption is justified. This result is given by the fact that the spectrum curvature is helpful to better describe the pion bump which is below $1\, \rm GeV$. Furthermore, the Fermi-LAT sensitivity degrades above $10\, \rm GeV$, leading to signal models being degenerate. This was also emphasised by \cite{Ambrosone:2022fip} which highlighted the importance of the upcoming CTA to discriminate between different spectral assumptions for local SFGs and SBGs.  

\begin{figure*}[t!]
    \centering
    \includegraphics[width=0.8\columnwidth]{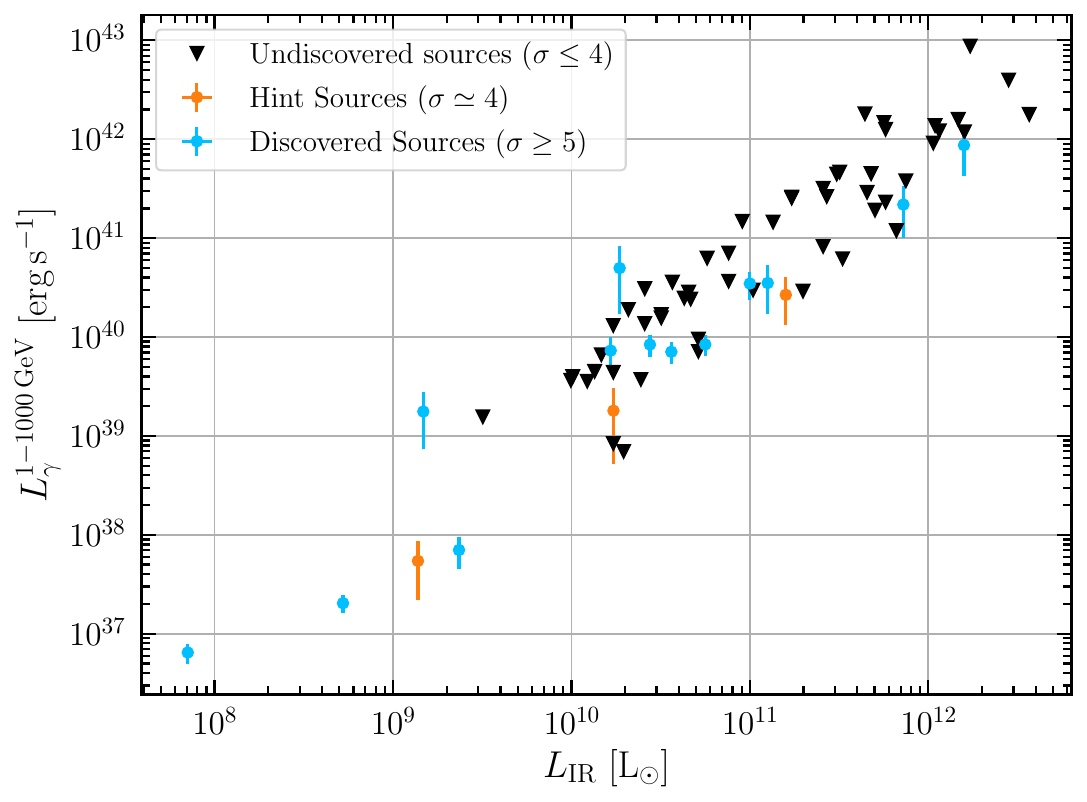}
    \caption{
   The $\gamma$-ray luminosity as a function of the total infrared luminosity for the entire sample. Cyan points represent discovered sources with a significance level of $\sigma \ge 5$, while orange points denote hint sources with a significance level of approximately $\sigma \simeq 4$. In both cases, the best fit scenarios along with their $1\sigma$ uncertainty are considered. For undiscovered sources ($\sigma \le 4$), represented by black triangles, we provide $95\%$ confidence level upper limits, fixing $\gamma=2.3$.}
    \label{fig:correlation_Lgamma_Lir}
\end{figure*}

For all the sources, we compute the $\gamma$-ray luminosity $L_{\gamma}$ between $1-1000\,\rm GeV$, using
\begin{equation}
    L^{1-1000~{\rm GeV}}_{\gamma} = 4\pi \frac{D^{2}_{L}(z)}{(1+z)^{2-\gamma}}F_{1-1000~{\rm GeV}}
\end{equation}
where
\begin{equation}
    F_{1-1000~{\rm GeV}} = \int_{1\, \rm GeV}^{1000\, \rm GeV} E\,\frac{{\rm d}F}{{\rm d}E} \, {\rm d}E
\end{equation}
is the integration of the differential flux measured weighted by the energy, and $z$ is the redshift of the source, directly related to the luminosity distance $D_{L}(z)$. Fig.~\ref{fig:correlation_Lgamma_Lir} shows the $L_{\gamma}$ in the energy range $[1,\,1000]~{\rm GeV}$ versus the $L_{\rm IR}$ for the samples A and B. We report the best-fit values and the corresponding $1\sigma$ uncertainty for all the discovered sources as well as for the three sources which give us a $4\sigma$ hint of emission. On the other hand, for the undiscovered sources, we report the $95\%$ CL upper limit assuming a $E^{-2.3}$ spectrum. In the plot, we also take into account a $10\%$ uncertainty in each distance and $5\%$ in $L_{\rm IR}$ as reported by~\cite{Ajello:2020zna}.

\section{On the Non-thermal emission from SFGs and SBGs}\label{sec:theoretical_model}

The results presented in the previous section have important repercussion on the CR transport mechanisms occurring inside these sources. Indeed, since photons produced by hadronic interactions usually carry $10\%$ of the parent energy of CRs, the $\gamma$-ray spectra are expected to inherit the properties of the CR distribution inside the sources. In order to assess such implications, we use a model describing the non-thermal emission of the sources. In general, since we expect the emission of SBGs to be dominated by their nuclei, we can neglect the spatial dependence of the CR diffusion. Hence, we can study the CR transport under the leaky-box model equation where the CR transport is modelled by a balance among different competing processes: the injection term of the sources such as SNRs, the escape phenomena (advection and diffusion) and the energy-loss mechanism such as hadronic collisions:
\begin{equation}\label{eq:leaky}
    \frac{N_{\rm CR}(E)}{\rm \tau_{\rm esc}} + \frac{{\rm d}}{{\rm d}\rm E}\left[\frac{{\rm d}E}{{\rm d}t}\cdot N_{\rm CR} (E)\right] = Q(E)
\end{equation}
where ${\rm d}E/{\rm d}t = -E/\tau_{\rm loss}$, with $\tau_{\rm loss}$ being the energy-loss timescale, $\tau_{\rm esc}$ is the escape timescale, and $Q(E)$ is the injection spectrum of SNRs. We assume the injected spectrum to be a power-law with a $E_{\rm max} = 10\, \rm PeV$ exponential cut-off consistent with our previous results~\citep{Ambrosone:2020evo,Ambrosone:2022fip} and we neglect any other spectral feature of the injected spectrum (see below for further remarks about the chosen cut-off). The normalisation is set as
\begin{equation}
    \int_{m_p}^{+\infty} Q(E)\, E \, {\rm d}E = \eta \, E_{\rm SN}\, R_{\rm SN}
\end{equation}
Hence, the total energy injected into CRs is $\eta = 10\%$ of the total $E_{\rm SN}=10^{51}\, \rm erg$ emitted by SNRs. The quantity $R_{\rm SN}$ is the SNRs rate which is expected to be tightly connected to the infrared luminosity according to the empirical relation~\citep{Kornecki:2021xiy,2003ApJ...586..794B} 
\begin{equation}\label{eq:Rsn_SFR}
    R_{\rm SN} = \frac{1}{83} 1.36\cdot 10^{-10} \left(\frac{L_{\rm IR}}{L_{\odot}}\right) \left(1+ \sqrt{\frac{10^9\,L_{\odot}}{L_{\rm IR}}}\right)\,{\rm yr^{-1}}
\end{equation}
which takes advantage of the Chabrier Initial mass function (IMF), consistent with $83\, M_{\odot}$, converted in new stars for each supernova explosion. In other words, the SFR is connected to $R_{\rm SN}$ through ${\rm SFR} / (M_{\odot}\, {\rm yr^{-1}}) = 83 \, M_{\odot}\,  R_{\rm SN} /(\rm yr^{-1})$. We emphasise that Eq.~\ref{eq:Rsn_SFR} is not linear because the infrared luminosity itself is not a perfect tracer of the SFR.

In general, the solution to Eq.~\ref{eq:leaky} can be approximated as~\cite{Kornecki:2021xiy}
\begin{equation}
   N_{\rm CR} (E) = \frac{\tau_{\rm tot}(E)}{E} \int_{E}^{+\infty} Q(E') {\rm d}E' \simeq  \frac{\tau_{\rm tot}(E) \, Q(E)}{\gamma -1}
\end{equation}
where $\tau_{\rm tot} = (\tau_{\rm loss}^{-1} + \tau_{\rm esc}^{-1})^{-1}$ and the last passage holds for $Q(E) \propto E^{-\gamma}$. For SBGs, pp interactions should be the dominant CR energy-loss mechanism. Therefore, $\tau_{\rm loss} = \tau_{\rm pp}$.  In turn, the escape timescale is given by the competition between CR advection and diffusion phenomena. While it is expected that their relative contribution to change across the whole SFR range~$(10^{-2}-10^{3}\, \rm M_{\odot}\, \rm yr^{-1})$~\cite{Werhahn:2021bal,Werhahn:2021jvy,Werhahn:2023osl,Pfrommer:2017jau}, these timescales are strongly model-dependent as well as dependent on the assumption for their scaling relation with the SFR. Indeed, although Refs.~\cite{Peretti:2018tmo,Peretti:2019vsj,Ambrosone:2020evo,Ambrosone:2021aaw,Ambrosone:2022fip} have shown advection to be important as escape phenomenon for SFGs and SBGs, Refs.~\cite{Krumholz:2019uom,Roth:2021lvk,Roth:2022hxc} have argued that advection should be suppressed in interstellar medium (ISM) ambient in SBGs, proposing a major role played by diffusion phenomena.  Furthermore, whereas Refs.~\cite{Peretti:2018tmo,Peretti:2019vsj,Ambrosone:2020evo,Ambrosone:2021aaw,Ambrosone:2022fip} have modelled the diffusion coefficient using quasi linear theory assuming a pre-existent magnetic field turbulence, Refs.~\cite{Krumholz:2019uom,Roth:2021lvk,Roth:2022hxc} have  used self-generated diffusion from streaming instability. Given that it is not possible to distinguish between these scenarios with Fermi-LAT data~(see previous section), here we introduce an overall parameter-$F_{\rm cal}$- defined as
\begin{equation}\label{eq:frac_CR}
    N_{\rm CR} (E) =F_{\rm cal} \cdot \frac{\tau_{\rm loss}}{E} \int_{E}^{+\infty} Q(E') {\rm d}E' \simeq  \frac{F_{\rm cal} \, \tau_{\rm loss} \, Q(E)}{\gamma -1}
\end{equation}
in order to test if the $\gamma$-ray measurements of sample A and sample B might be interpreted in terms of star-forming activity. $F_{\rm cal}$ is defined between 0 and 1 and it can be interpreted as an average  fraction of CRs  between $10 \leq E_{\rm CR}/{\rm GeV} \leq 10^4$ actually losing their energy onto pp collisions producing $\gamma$-rays and neutrinos. A very small $F_{\rm cal}$ value  would correspond to a very strict constraint on the ability to confine high-energy protons by the source. $F_{\rm cal}$ can be expressed as
\begin{equation}\label{eq:fcal_model}
   F_{\rm cal} = \frac{\tau_{\rm esc}}{\tau_{\rm loss}} \bigg(\frac{\tau_{\rm esc}}{\tau_{\rm loss}}+1\bigg)^{-1} 
\end{equation}
For the following analysis, we assume $F_{\rm cal}$ to be constant, which allows us to estimate it directly from the $\gamma$-ray data without any  assumption on the magnetic field, gas density, wind velocity and energy dependence of the diffusion coefficients of the sources because we treat it as an effective number for each of the source in the sample. However, this restricts our analysis to assume that $\tau_{\rm esc}$ is only mildly energy-dependent in the whole SFR range analysed, leading to negligible diffusion phenomena. This might slightly overestimate  $F_{\rm cal}$ for low-SFR sources, where the role of diffusion might be more relevant~\cite{Kornecki:2021xiy}. However, we stress that all the SFGs discovered, from SMC to ARP 220, show  the same spectral behaviour~$(\sim E^{-2.3-2.4})$  totally consistent with the injected spectrum inferred for the Milky-way~\cite{Evoli:2019wwu,Caprioli:2020spz}. Therefore, an energy-independent escape timescale cannot be, at the moment, completely ruled out. 
From Eq. \ref{eq:frac_CR}, we can quantify the photon production rate following the analytical procedure of~\cite{Kelner:2006tc} (see also \cite{Kornecki:2021xiy}). For $E_{\gamma} > 100\, \rm GeV$, we have 
\begin{equation}\label{eq:kelner_photons} 
    Q_{\gamma}(E_{\gamma}) = c\,n_{\rm gas} \int_{\rm x_{\rm min}}^{1} \sigma_{pp} \left(\frac{E_{\gamma}}{x}\right) N_{\rm CR}\left(\frac{E_{\gamma}}{x}\right) \tilde{F}_{\gamma}\left(x,\frac{E_{\gamma}}{x}\right) \frac{{\rm d}x}{x}
\end{equation}
where $\tilde{F}_{\gamma}\left(x,{E_{\gamma}}/{x}\right)$ is defined in~\citep{Kelner:2006tc} (see Eqs. (58-61)) and $x_{\rm min} = 10^{-3}$. For lower energies, we can assume that the pions produced by pp collisions take $K_{\pi} = 17\%$ of the kinetic energy of the parent high-energy proton (delta-function approximation), having
\begin{equation}\label{eq:delta_function}
 \begin{split}
     Q_{\gamma}(E_{\gamma}) & = \frac{2c\,n_{\rm gas}}{K_{\pi}} \int_{E_{\gamma}+m_{\pi}^2c^4/E_{\gamma}}^{+\infty} \frac{1}{\sqrt{E_{\pi}^2 -m_{\pi}^2}}\sigma_{pp} \left(m_p c^2 + \frac{E_{\pi}}{K_{\pi}}\right) \\
     & \times \, N_{\rm CR}  \left(m_p c^2 + \frac{E_{\pi}}{K_{\pi}}\right) \,{\rm d}E_{\pi}
 \end{split}
\end{equation}
At $E_{\gamma} = 100\, \rm GeV$, Eq.~\eqref{eq:delta_function} is scaled in order to match Eq.~\eqref{eq:kelner_photons}. 
The final $\gamma$-ray flux at Earth is given by 
\begin{equation}\label{eq:flux_earth}
    \frac{{\rm d}F_{\gamma} (E_{\gamma},z)}{{\rm d}E} = \frac{(1+z)^2}{4\pi D_{L}(z)^2} Q(E_{\gamma}(1+z)) e^{-\tau(E_{\gamma},z)} 
\end{equation}
where $z$ is the redshift of the source, $D_{L}(z)$ is the luminosity distance, and $\tau(E,z)$ is the optical depth for photons travelling through EBL and CMB. For the computation of the opacity, we employ the model of~\cite{Franceschini:2017iwq}.
From pp interactions, we expect production of high-energy neutrinos as well and we estimate their flux using the same procedures as for $\gamma$-rays. In particular, for $E_{\nu} > 100\, \rm GeV$, we have that
\begin{equation}\label{eq:kelner_neutrinos}
\begin{split}
    Q(E_{\nu}) & = \frac{1}{3} c\,n_{\rm gas} \int_{\rm x_{\rm min}}^{1} \sigma_{pp} \left(\frac{E_{\nu}}{x}\right) N_{\rm CR}\left(\frac{E_{\nu}}{x}\right) \left[ \tilde{F}_{\nu^{1}_{\mu}}\left(x,\frac{E_{\nu}}{x}\right) + \right. \\
    & \left.+ \tilde{F}_{\nu^{2}_{\mu}}\left(x,\frac{E_{\nu}} {x}\right)+\tilde{F}_{\nu_{e}}\left(x,\frac{E_{\nu}}{x}\right)\right] \,\frac{{\rm d}x}{x}
    \end{split}
\end{equation}
where $\tilde{F}_{\nu^{1}_{\mu}}\left(x,E_{\nu}/x\right)$, $\tilde{F}_{\nu^{2}_{\mu}}\left(x,E_{\nu}/x\right)$ and $\tilde{F}_{\nu_{e}}\left(x,E_{\nu}/x\right)$ take into account all the neutrinos produced in the interactions and are defined by \cite{Kelner:2006tc}. The factor 1/3 is due to the fact that we expect an equal flavour ratio at Earth. The final neutrino flux is given by
\begin{equation}\label{eq:flux_earth_neutrinos}
\frac{{\rm d} F_{\nu} (E_{\nu},z)}{{\rm d}E} = \frac{(1+z)^2}{4\pi 
 D_{L}(z)^2} Q_{\nu}(E_{\nu}(1+z))
\end{equation}
Before concluding this section, we emphasise that the $\gamma$-ray spectra of SFGs and SBGs might be contaminated also by leptonic contributions such as Inverse compton and Bremsstrahlung as well as by the AGN related activity hosted by some of the sources in sample A and B~\cite{Inoue:2022yak,Liu:2017bjr,Senno:2015tra,Peretti:2023xqk}~(see app. \ref{app:AGN_contamination}). Therefore, our results might be relatively interpreted as upper limits for $F_{\rm cal}$ which corresponds to conservative constraints on the star-forming activity of the sources. Regarding the leptonic contributions, in our approximation where diffusion is negligible, the contribution from leptonic photons is expected very limited~\cite{Ambrosone:2020evo} above $1\,\rm GeV$. However, this intrinsically assume a proton to primary electron ratio of $K_{ep} = 50$ consistently with the Milky-way. Indeed, lower values would lead to a major role for primary electrons since they are usually trapped in the SFG environments cooling down much faster than protons~\cite{Peretti:2018tmo}.

\section{On the correlation between gamma-rays and star-forming activity}\label{sec:correlation_Fcal_Rsn}

In this section, we discuss our constraints on the calorimetric fraction $F_{\rm cal}$ from $\gamma$-ray observations and its correlation with $R_{\rm SN}$. Previous studies~\citep{2012ApJ...755..164A,Ajello:2020zna,Rojas-Bravo:2016val,Xiang:2023dww} have tested the relation $F_{\rm cal} \simeq A \left({R_{\rm SN}}/{1\, \rm yr^{-1}}\right)^{\beta}$. However, this relation cannot be valid for a wide SFR range $[10^{-3}-10^{3}]\, M_{\odot}\, \rm yr^{-1}$, since the calorimetric limit cannot be exceeded. 
In order to test a physically motivated relation between $F_{\rm cal}$ and $R_{\rm SN}$, we exploit the fact that  $\tau_{pp} = (k \cdot n_{\rm gas} \sigma_{pp} c)^{-1}$ where $k = 0.5$ is the mean inelasticity of the process and $\tau_{\rm esc} = H/v_{\rm wind}$ with $H$ being the height of the nucleus  and $v_{\rm wind}$ is the velocity of the galactic winds. Both $n_{\rm gas}$ and $v_{\rm wind}$ are expected to scale with $R_{\rm SN}$. Indeed, according to the kennicutt relation~\cite{Kennicutt:1998zb,2021ApJ...908...61K}, we have a strict connection between $n_{\rm gas}$ and $R_{\rm SN}$, namely $n_{\rm gas} \propto R_{\rm SN}^{2/3}$~\cite{Ambrosone:2021aaw,Ambrosone:2022fip}. By contrast, the wind velocities have been found to correlate with the SFR as~$v_{\rm wind} \propto R_{\rm SN}^{0.15-0.30}$~\cite{2018Galax...6..138R}. All of this leads to $\tau_{\rm esc}/\tau_{\rm pp} \sim A R_{\rm SN}^{0.30-0.50}$.  
Therefore, in the present paper we probe the following relation between $F_{\rm cal}$ and $R_{\rm SN}$
\begin{equation}\label{eq:fcal_fit}
   F_{\rm cal} = A \left(\frac{R_{\rm SN}}{\rm yr^{-1}}\right)^{\beta}\,\left(1 +  A \left(\frac{R_{\rm SN}}{\rm yr^{-1}}\right)^{\beta}\right)^{-1}
\end{equation}
with $A$ and $\beta$ free parameters to be deduced from data. 

We notice that, for small value of $R_{\rm SN}$, Eq.~\eqref{eq:fcal_fit} becomes consistent with a pure power-law relation as tested by previous study~(in appendix \ref{app:power-law}, we discuss also the power-law fit).

In order to test Eq.~\eqref{eq:fcal_fit}, for each source we estimate $R_{\rm SN}$ from the infrared luminosity according to Eq.~\eqref{eq:Rsn_SFR} and we calculate the calorimetric fraction as described in Eq.~\eqref{eq:frac_CR} by matching the measured integrated spectrum with the theoretical one using the model described in the previous section.

For the discovered sources (sample B), we evaluate the best-fit scenario and the $\pm 1\, \sigma$ values. For the undiscovered sources (sample A), we utilise the best-fit scenario for the fixed $\gamma = 2.3$ and for the uncertainty, we consider the difference between $F^{95\% \rm CL}$ and $F^{\rm best}$.

In addition to the statistical uncertainties inferred by Fermi-LAT data, we also take into account the systematic uncertainties affecting $ F_{\rm cal}$. On this regard, uncertainties on the source distance and rate of supernovae explosions as well as the detector systematics play a crucial role. As we mentioned above, the distance and the infrared luminosity provide an uncertainty of the order of $10\%$ and $5\%$, respectively. By contrast, the uncertainty on $R_{\rm SN}$ might also come from the IMF and the amount of mass converted in new star from each supernova explosions. The total uncertainty on $R_{\rm SN}$ is difficult to reliably assess and it may vary within $20-40\%$ (see \cite{2012ApJ...755..164A} for further details). For the following discussion, we consider a systematic uncertainty of $20\%$ on our estimates of $R_{\rm SN}$ and in the appendix~\ref{app:Rsn_systematic_uncertainty} we discuss the impact of a higher uncertainty. Regarding the detector systematic uncertainty, we consider a conservative uncertainty of $10\%$.\footnote{see \url{https://fermi.gsfc.nasa.gov/ssc/data/analysis/scitools/Aeff_Systematics.html} for more details.} Summing all the systematic uncertainties in quadrature, we obtain an overall $30\%$ uncertainty on each value of $F_{\rm cal}$.
\begin{figure}[t!]
    \centering
\includegraphics[width=0.8\columnwidth]{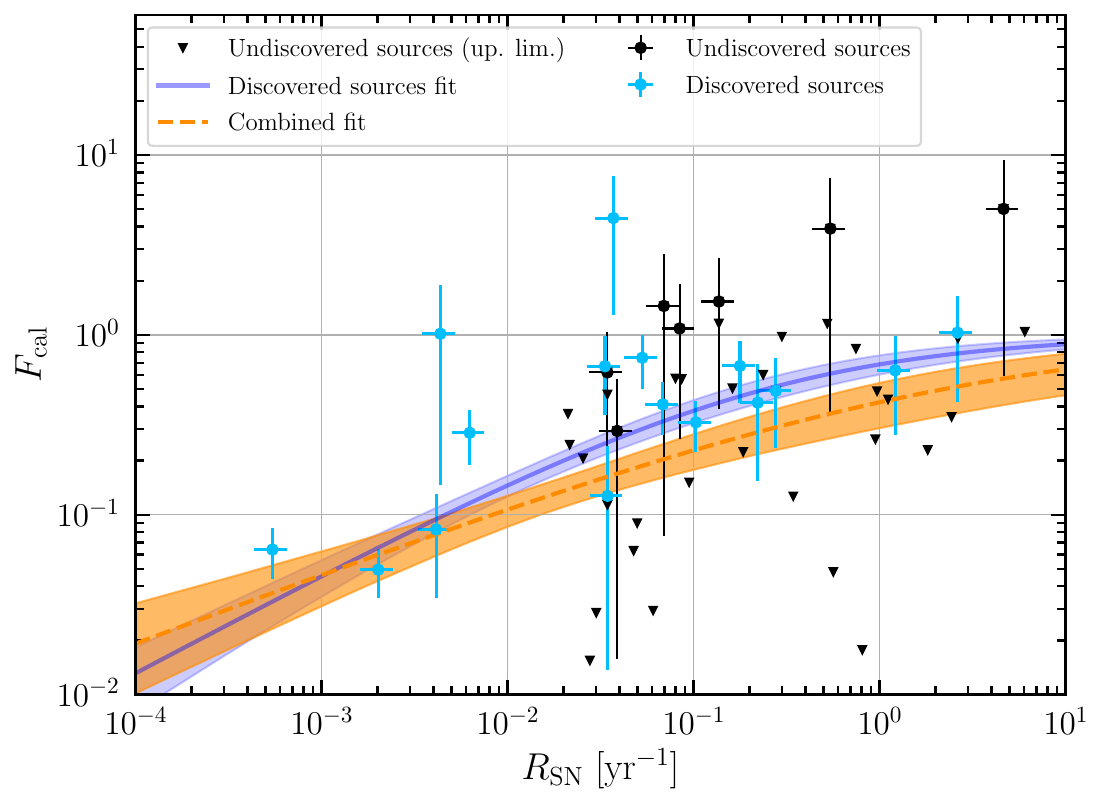}
    \caption{$F_{\rm cal}$ in terms of $R_{\rm SN}$ for the whole sample. Cyan points denote discovered sources, whereas black points denote undiscovered sources. Specifically, for sources exhibiting a flux compatible with zero, we present 95\% CL upper limits indicated by black triangles. For all the sources,  $F_{\rm cal}$  and $R_{\rm SN}$values are reported respectively with 30\% and 20\% systematic uncertainties. We also report the best-fit and the corresponding $1\sigma$ uncertainty band of the fit performed over the whole sample (orange) and over discovered sources (blue). }
    \label{fig:Fcal_vs_Rsn}
\end{figure}

Fig.~\ref{fig:Fcal_vs_Rsn} shows the obtained $F_{\rm cal}$ both for undetected and detected sources as a function of $R_{\rm SN}$, as well as the $1\sigma$ bands from the fit of Eq.~\eqref{eq:fcal_fit} according to two different samples of galaxies:
\begin{itemize}
    \item \textbf{Discovered sources}, for which we find $A = 2.2 \pm 0.8 $ and $\beta = 0.55 \pm 0.08$;
    \item \textbf{Combined sources}, namely discovered + undiscovered sources, for which we find $A = 0.7_{-0.2}^{+0.3}$ and $\beta = 0.39\pm 0.07$.
\end{itemize}
Interestingly, even though the undiscovered sources are characterised by higher uncertainties, they are anyway able to constrain the $F_{\rm cal}$ fit especially in the range $[0.01-1]\,\rm yr^{-1}$. In the lowest range for $R_{\rm SN}$, the fit is totally dominated by the galaxies of local group (SMC, LMC, M 31 and M 33). On this regard, we have verified, in app.~\ref{app:AGN_contamination}, that removing M31 from the fit (because of its soft spectrum) does not impact our results. 
We a posteriori notice that our results are completely in agreement with the expected scaling values for CR escape phenomena dominated by advection. We have also verified that our results for low SFRs are consistent with the $\gamma$-ray measurements of the central part of our galaxy, the central molecular zone (CMZ). In fact, the Fermi-LAT and the HESS collaborations have reported  $\sim E^{-2.4}$ spectra for the CMZ~\cite{Marinelli:2018lzs,HESS:2017tce}. In particular, assuming the observed $R_{\rm SN} = 2\cdot 10^{-4}\, \rm yr^{-1}$ \cite{2023MNRAS.518.6273S} for the CMZ and the corresponding $F_{\rm cal}$ from our combined fit, we obtain that our predicted $\gamma$-ray flux is consistent with the diffuse measurements from the galactic ridge of the CMZ.
We mention that a greater component coming from leptonic processes would lead to a smaller calorimetric fraction potentially constraining even more the properties of the population. We can extract that $F_{\rm cal} \gtrsim 50\%$ for ${\rm SFR} \gtrsim 190_{-130}^{+1230}\, M_{\odot}\, \rm yr^{-1}$ when considering the whole sample. This might reduce the degree of calorimetry of Ultra Luminous Infrared Galaxies (ULIRGs) (sources with ${\rm SFR} \gtrsim 100\, M_{\odot}\, \rm yr^{-1}$), 
although, this information at the moment is mainly driven by galaxies with lower IR luminosity, since Fermi-LAT is not yet sensitive enough to directly probe the calorimetric scenario within ULIRGs, due to their large distances. We highlight that MHD simulations~(e.g. \cite{Werhahn:2021bal}) have theoretically predicted that calorimetric limit~$(F_{\rm cal} \simeq 1)$ cannot be reached by SBGs, although this conclusion is driven by an assumed diffusion coefficient of $3\cdot 10^{28}\, \rm cm^2\, \rm s^{-1}$ at 3 GeV, which is higher than expected in extreme environments such as ULIRGs~\cite{Peretti:2018tmo}. On the contrary, our results are entirely driven by the latest data, making them the most current and robust constraints. 


\section{Extrapolation to the diffuse emissions}\label{sec:diffuse}

We can use the calorimetric fraction of local SFGs and SBGs evaluated in the previous section to constrain the diffuse non-thermal emission of the entire source population. The diffuse emission, per solid angle, is given by 
\begin{equation}\label{eq:diffuse}
   \begin{split}
     \phi^{\rm diff}_{\gamma, \nu} = & \frac{c}{4\pi H_{0}} \int_{0}^{z_{\rm max}} \frac{{\rm d}z}{E(z)} \int_{10^6 L_{\odot}}^{\infty} \frac{{\rm d}L_{\rm IR}}{\ln (10)\,L_{\rm IR}} \mathcal{S}_{\rm SFR}(L_{\rm IR},z)\\
    & \times Q_{\gamma, \nu}\left(E(1+z), \rm R_{\rm SN}(L_{\rm IR}), F_{\rm cal} (R_{\rm SN}(L_{\rm IR}))\right)\,  e^{-\tau_{\gamma,\nu}(E,z,L_{\rm IR})}
 \end{split}
\end{equation}
where z is the redshift, $E(z) = \sqrt{\Omega_{M}(1+z)^3 + \Omega_{\Lambda}}$, $\mathcal{S}_{\rm SFR}(L_{\rm IR},z)$ is the density of the sources as a function of the infrared luminosity,  $Q_{\gamma, \nu}$ are the $\gamma$ and neutrino production rate for each source, and $\tau_{\nu} = 0$ and $\tau_{\gamma}(E,z,L_{\rm IR})$ accounts for the CMB+EBL absorption of photons as well as for internal absorption phenomena~\citep{Ambrosone:2021aaw}. We highlight that in Eq.~\ref{eq:diffuse} we use $L_{\rm IR} = 10^6 L_{\odot}$ as a lower limit for the infrared luminosity corresponding at ${\rm SFR }\sim 4\cdot 10^{-3}\, M_{\odot}\, \rm yr^{-1}$. Increasing such a value to $10^{10}\, L_{\odot} (\sim 1\, M_{\odot}\, \rm yr^{-1})$ results in a reduction of the flux by $\sim$$5\%$ only, since the bulk of the emission comes from sources with higher star formation rates. For the density of the sources, we use the approach described by~\cite{Kim:2023xxh}, who have recently updated the distribution of the cosmic SFR using also JWST data. The distribution is given in terms of a Schechter function
\begin{equation}\label{eq:Schechter}
    \mathcal{S}_{\rm SFR}(L_{\rm IR},z) = \Phi^{*}(z) \left(\frac{L_{\rm IR}}{L^{*}(z)}\right)^{1-\alpha} e^{-\frac{1}{2\sigma^2}\log_{10}^{2}\left(1+\frac{L_{\rm IR}}{L^{*}(z)}\right)}
\end{equation}
which behaves as a power-law for $L_{\rm IR} \ll L^{*}(z)$ and as a Gaussian in $\log_{10}(L_{\rm IR})$ for $L_{\rm IR} \gg L^{*}(z)$.
The redshift parameter evolutions are not simply set by power-laws, but rather follow skew Gaussian distributions~\cite{2011MNRAS.416...70G}
\begin{eqnarray}\label{eq:lstar}
    \log_{10}(L^{*}(z)) &=& \log_{10} (L^{*}(0)) + \frac{A_L}{2\pi\omega} e^{-z^2/(2\omega^2)} \, {\rm erf}\left(\frac{k \, z}{\omega}\right) \\
    \label{eq:phistar}
    \log_{10}(\Phi^{*}(z)) &=& \log_{10} (\Phi^{*}(0)) + \frac{A_{\Phi}}{2\pi\omega} e^{-z^2/(2\omega^2)} \, {\rm erf}\left(\frac{k \, z}{\omega}\right)
\end{eqnarray}
where $\text{erf}(x)$ is the error function, $k$ is called the shape parameter, $\omega$ the scale factor, $A_L$ and $A_{\Phi}$ are the normalisation for the evolution of $L^{*}(z)$ and $\Phi^{*}(z)$, respectively. Eqs.~\eqref{eq:lstar} and~\eqref{eq:phistar} provide physical representations of the evolution, allowing for different peaking redshifts as well as asymmetric increasing/decreasing rates for several populations~\cite{2011MNRAS.416...70G}. In fact, one of the main advantages of such a parameterisation is that it can be divided for distinct source classes. Here, we consider SFGs and SBGs, taking  the values reported in Tab.~\ref{tab:Schechter}. They provide excellent agreement with the ones reported by~\cite{Kim:2023xxh} (see their Fig.~10). Some parameters are also in agreement with the ones reported by~\cite{2011MNRAS.416...70G}.
\begin{table}[h!]
    \centering
    \begin{tabular}{c|c|c|c|c|c|c|c|c|}
       Source Class  & $A_{L}$ & $A_{\Phi}$ & $k$ & $\omega$ & \small{$\Phi^{*}(0)\,  [\rm Mpc^{-3}\, \rm dex^{-1}]$} & $L^{*}(0) \, [L_{\odot}]$ & $\alpha$ & $\sigma$ \\
        SFGs & 1.01   &  3.79  & 5.11  & 2.40  &  $1.2\cdot 10^{-3}$ & $3.2\cdot 10^{10}$ & 1,35 & 0.300 \\
        SBGs & 11.95    & 8.50  &  3.10 &  3.50  & $3\cdot 10^{-5}$ & $1.5\cdot 10^{10}$ & 0.05 & 0.465 \\
    \end{tabular}
    \caption{Parameters defining the Schechter function in Eqs.~\eqref{eq:Schechter},~\eqref{eq:lstar} and~\eqref{eq:phistar}. The first four parameters (from left to right) are the ones reported by~\cite{Kim:2023xxh}, while the remaining parameters are obtained to match the SFG and SBG distributions reported in their Fig.~10.}
    \label{tab:Schechter}
\end{table}
Finally, for $F_{\rm cal} (L_{\rm IR})$ we use Eq.~\eqref{eq:fcal_fit} with parameters inferred by the data of both the discovered sources and the total sample (discovered + undiscovered) sources.
\begin{figure*}[t!]
    \centering
    \includegraphics[width=0.49\columnwidth]{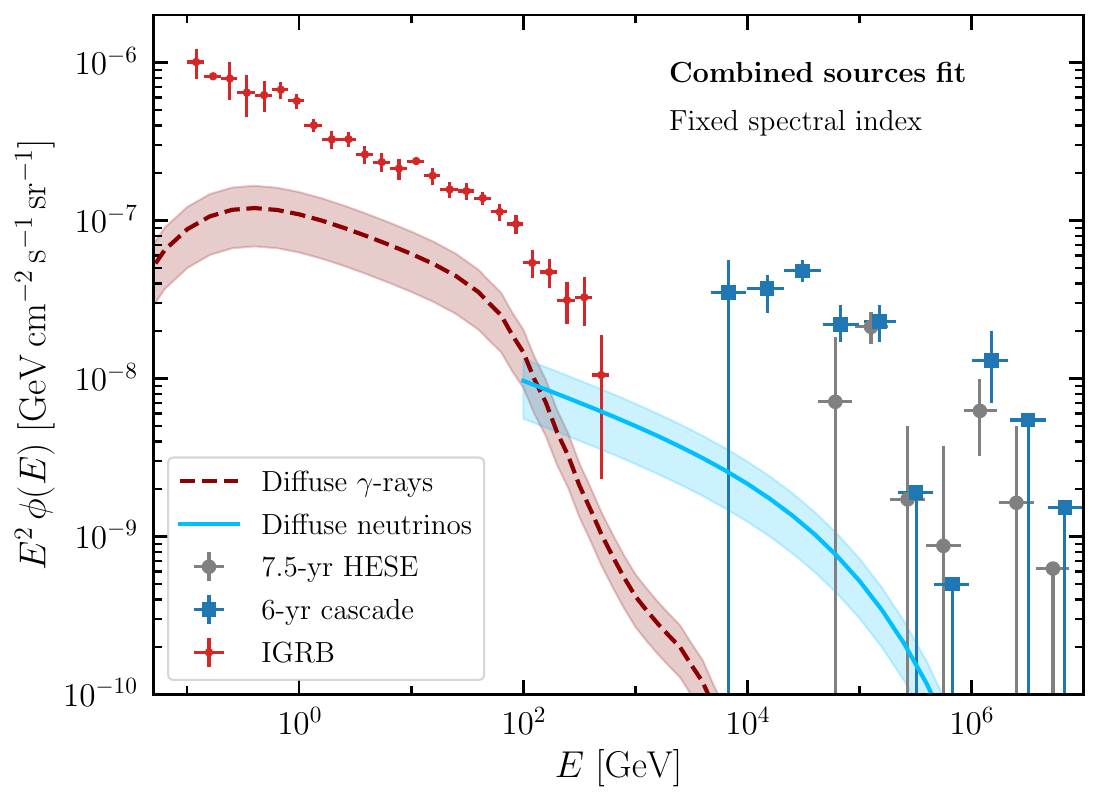}
    \includegraphics[width=0.49\columnwidth]{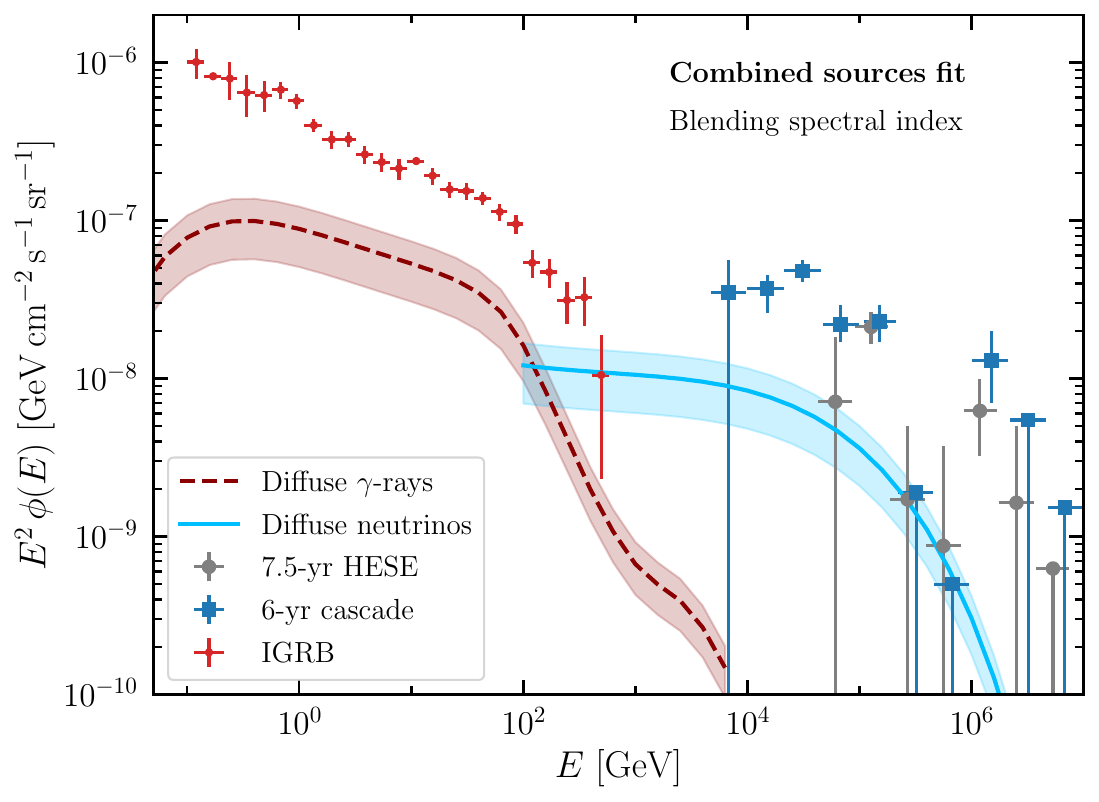}
    \caption{\label{fig:diffuse_total}
    Left: Diffuse $2\sigma$ $\gamma$-ray (dark red) and neutrino (cyan)  bands predicted with the fit over the whole sample. On the left, the spectral index is fixed at $2.3$ for each source. Right: the same  but considering a spectral index blending. In both panels, the fluxes are compared with the Isotropic Gamma-Ray Background (IGRB) measured by Fermi-LAT \cite{Fermi-LAT:2014ryh}, the 6 year Cascade neutrino flux \cite{IceCube:2020acn} and 7.5 year HESE data \cite{IceCube:2020wum} measured by the IceCube neutrino Observatory.}
\end{figure*}

Fig.~\ref{fig:diffuse_total} shows the final $\gamma$-ray (in dark red colour) and neutrino (in cyan colour) fluxes for the combined fit including discovered and undiscovered sources. 

On the left panel, we have fixed spectral index to $\gamma = 2.3$, while on the right we have used a spectral index distribution (blending scenario) provided by a superposition  of Gaussian distributions with mean values equal to the best-fit spectral index for discovered sources and with standard deviation equal to their corresponding uncertainty~(for the spectral index blending flux calculation, we employ the same technique as in \cite{Ambrosone:2020evo}). In this approach, the injected spectral index follows a continous distribution which allows also for spectral indexes lower than 2. The theoretical predictions are compared with the Isotropic Gamma-Ray Background (IGRB) measured by Fermi-LAT~\citep{Fermi-LAT:2014ryh}, the 6-year cascade neutrino flux~\citep{IceCube:2020acn} and 7.5-year HESE data~\citep{IceCube:2020wum} measured by the IceCube neutrino Observatory. The fluxes are dominated by distant sources with a contribution peaking at $z \simeq 1$. Furthermore, the bulk of the emissions come from ULIRGs saturating almost 51\% of the emissions (see the appendix~\ref{app:diffuse_flux} for details).
We find that the total contribution to the extra-galactic gamma-ray background (EBG)~\citep{Fermi-LAT:2014ryh} between 50 GeV and 2 TeV is $\simeq (12\pm 3)\%$, almost independent on the spectral index distribution considered. The neutrino spectrum, on the contrary, is strongly dependent on the spectral index distribution. Indeed, fixing a spectral index $\gamma = 2.3$ provides a soft diffuse spectrum which can explain only $(4^{+1}_{-2})\%$ of the 6-year cascade flux between 10 TeV and 1 PeV. By contrast, the spectral index blending hardens the spectrum and allows for the neutrino spectrum to explain $(18^{+3}_{-5})\%$ of the 6-year cascade IceCube flux.
This result is mainly driven by  sources with $\gamma \lesssim 2$ which contaminate the overall distribution of $10\%$. Indeed, if we only considered the distribution  with $\gamma \ge 2$, the neutrino spectrum would be at level of $\sim 7\%$ of the 6-year cascade IceCube flux, reducing the observable signature of the spectral index blending. We notice that, at the moment, some observed $\gamma$-ray spectra of young SNRs might point to very hard injected proton spectra~\cite{Zeng:2021mhf,Morlino:2017gck}, although it is still controversial if this is a true signature given by hard hadronic spectra or leptonic processess. We also underline that, given the limited number of discovered sources, it is not possible to derive a robust distribution for the spectral indexes and its impact might vary also with respect to the the statistical treatment of the data~\cite{Ambrosone:2020evo}.
The neutrino flux is also sensitive on the chosen high-energy cut-off for CRs. Indeed, Ref.~\cite{Peretti:2019vsj} has argued that the highly dense environment of SBGs might cause turbulent amplification of the   magnetic field, leading to an $E_{\rm max} \simeq 50-100\, \rm PeV$. Furthermore, it is possible that since $E_{\rm max}$ is correlated to the magnetic field value in SBGs, it might have a non-trial dependence on the SFR leading to a further signature, which we leave for future explorations. 

Our results are completely consistent with previous works~\cite{2012ApJ...755..164A,Ajello:2020zna} which employ the same technique and also with our previous multi-messenger analysis~\cite{Ambrosone:2020evo}. We find that SFGs and SBGs contribute significantly less to the EGB than the limits imposed on non-blazar sources~\cite{Bechtol:2015uqb},  which has sensibly reduced the possible role of  SFGs and SBGs to the EGB suggested by earlier works~\cite{Tamborra:2014xia}. 
On the other hand, we find a slightly lower contribution than~\cite{Peretti:2019vsj} due to several reasons. Firstly, we assume a steeper injected spectrum; secondly, the authors of Ref.~\cite{Peretti:2019vsj} have assumed the background photon energy density to be equal to the M82 value in order to estimate the contribution of internal absorption, while we take into account the fact that the energy density of background photon linearly scales with $R_{\rm SN}$~\cite{Ambrosone:2021aaw}. This leads to a further suppression of photons for high SFR sources. Thirdly, we assume a lower value for the high-energy cut-off for CRs. Moreover, there is a different assumption on $F_{\rm cal}$, since the authors have estimated only the contribution of high SFR sources assuming that they all had the same $F_{\rm cal}$ as M82 which is an assumption mainly tuned on discovered sources. On this regard, we assess 
\begin{figure*}[t!]
    \centering
    \includegraphics[width=0.49\columnwidth]{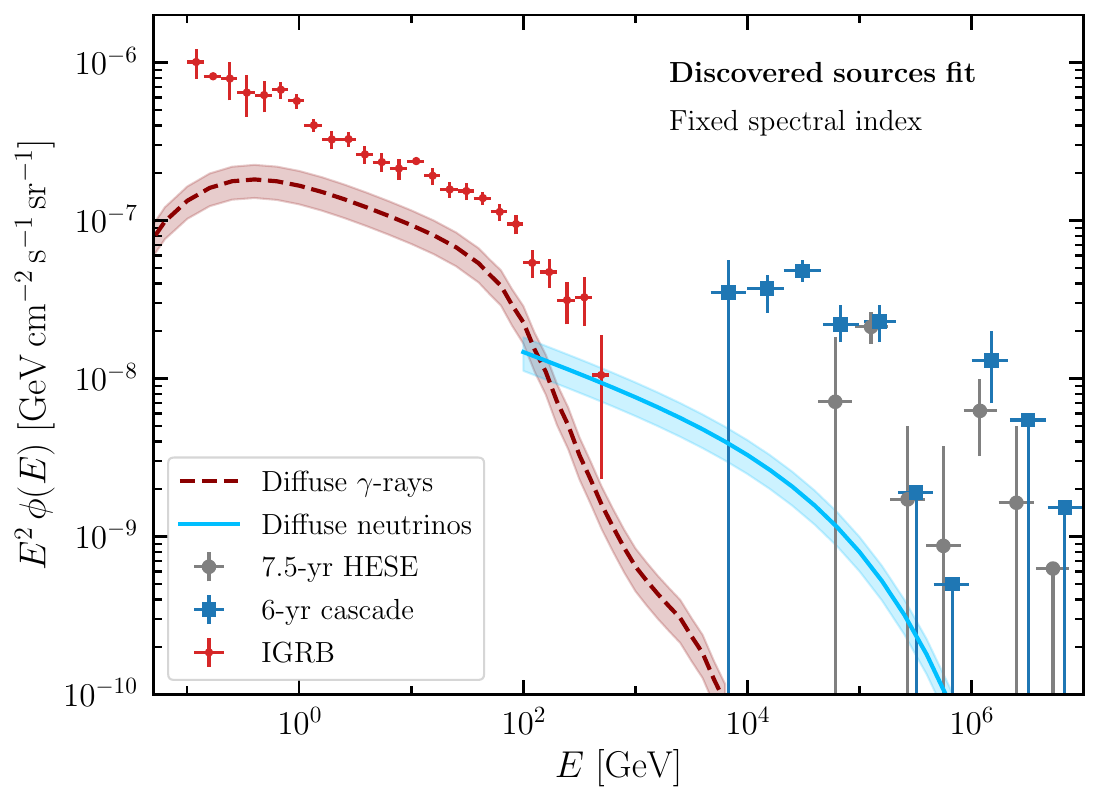}
    \includegraphics[width=0.49\columnwidth]{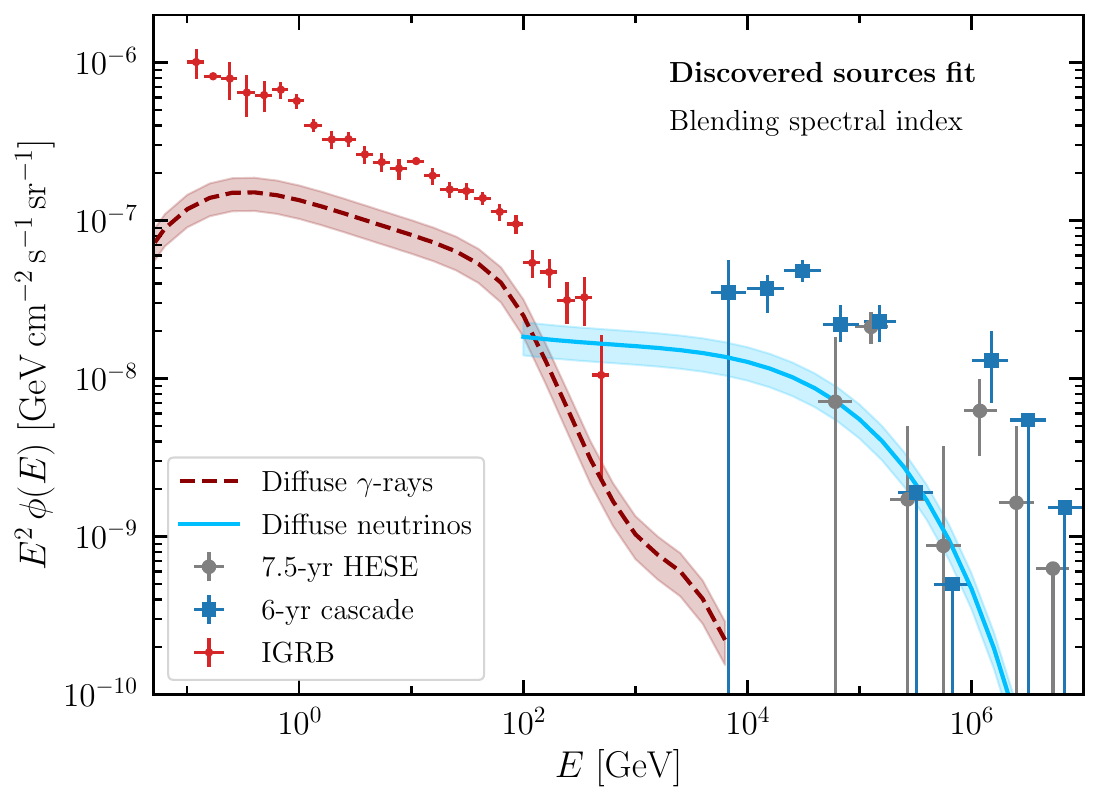}
    \caption{\label{fig:diffuse_scoperte}
    Left: Diffuse $2\sigma$ $\gamma$-ray (dark red) and neutrino (cyan)  bands predicted with the fit over discovered sources. On the left, the spectral index is fixed at $2.3$ for each source. Right: the same but considering  a spectral index blending. In both panels, the fluxes are compared with the Isotropic Gamma-Ray Background (IGRB) measured by Fermi-LAT \citep{Fermi-LAT:2014ryh}, the 6 year Cascade neutrino flux \citep{IceCube:2020acn} and 7.5 year HESE data \citep{IceCube:2020wum} measured by the IceCube neutrino Observatory.}
\end{figure*}
the impact of the undiscovered sources in the $F_{\rm cal}$, in Fig.~\ref{fig:diffuse_scoperte} where we show the diffuse $\gamma$-ray and neutrino spectra obtained with the fit of the discovered sources only. In this case, we obtain that SFGs and SBGs may contribute more to the EGB and the diffuse neutrino flux~($\sim$ a factor 2) compatible with the estimates given by~\cite{Peretti:2019vsj}. Therefore, undiscovered sources are not only important to correctly estimate the significance of the correlation between $F_{\rm cal}$ and $R_{\rm SN}$, but they are also necessary to correctly extrapolate information to the whole source population~\citep{2012ApJ...755..164A}. This is crucial because, typically, analyses which attempts to constrain the properties of SFGs and SBGs tune their models on the sources discovered in the $\gamma$-ray range, but there are a lot of sources with the similar astrophysical properties which have not been detected and they should be taken into account if the entire source population share the same properties.

\vspace{5pt}
However, SFGs and SBGs are still unable to completely saturate the IGRB between $\sim 1-100\, \rm GeV$ as recently obtained by~\cite{Roth:2021lvk}. Also in this case, the difference with our approach is given by several factors such as the source count and the CR transport model. Moreover, their assumed CR transport allows for $F_{\rm cal}$ being a function of redshift~(see  Fig. 4 in the extended data section in \cite{Roth:2021lvk} ). Indeed, if distant sources, which dominate the diffuse background (see~app. \ref{app:diffuse_flux}), are more intrinsically luminous, then SFGs and SBGs are allowed to explain a higher portion of the diffuse fluxes. We emphasise that in our approach Eq.~\eqref{eq:fcal_fit} is considered to be valid at each redshift even though only local sources has been used to constrain it. only future observation can challenge this assumption because at the moment Fermi-LAT is not sensitive enough to probe the calorimetric scenario for more distant sources. 
We point out that even though $F_{\rm cal}$ calculated with Fermi-LAT data corresponds to average values of $E_{\rm CR}$ between $10-10^4\, \rm GeV$, we extrapolate this calorimetric fraction also to higher energies in order to estimate the neutrino contribution. This, from one hand, it may be pessimistic since in case of energy-independent escape timescales, $F_{\rm cal}$ is logarithmically energy-increasing due to the energy behaviour of $\sigma_{pp}$. From the other hand,
at PeV energies, the diffusion process might not be negligible, leading to escape timescales might be energy dependent strongly suppressing the calorimetric fraction. On the whole, we find our approximation to be a reasonable trade-off, although it is difficult to quantify the uncertainty on the neutrino flux given the uncertainty on the nature of the diffusion process.

\section{Conclusions}\label{sec:conclusions}

In this paper, we have analysed 70 local sources, classified as star-forming and starburst galaxies, using 15 years of Fermi-LAT data. In order to reduce contamination from possible AGN activity as well as to reduce the possibility of mis-identification of sources from limited PSF, we have searched for photons with $E_\nu > 1\, \rm GeV$. We have found evidence at $\sim$$4\sigma$ for two nearby sources, M 83 and NGC 1365. On the contrary, even with 15 years of Fermi-LAT data, M33 still stands at $4\sigma$ due to an improved treatment of the background model.


We imposed strict upper limit at $95\%$ CL fixing a spectral index $\gamma = 2.3$ for the other sources. Exploiting these findings, we have then revisited the correlation between the $\gamma$-ray luminosity and the star formation rate for local star-forming and starburst galaxies. For the first time, we have studied this correlation under a physically-motivated relation between the calorimetric fraction and the rate of supernova explosions. We have found that there is a good agreement between the measurements and the theoretical model and that undiscovered sources play an important role in constraining the calorimetric fraction. This is crucial in order to capture the shared properties of these sources. 

Then, we have extrapolated this information to constrain the diffuse $\gamma$-ray and neutrino spectra of SFGs and SBGs, finding that they contribute about $12\%$ to the EGB above $50\, \rm GeV$. The corresponding neutrino flux is strongly dependent on the spectral index distribution along the source class. Indeed, if it is fixed at $\gamma = 2.3$ for the entire source spectrum, the contribution is negligible to the diffuse neutrino flux measured by ICeCube.

By contrast, if there is a continuous distribution of this parameter within the source class, the contribution to the diffuse neutrino flux could increase by up to 20\% because of sources with hard spectra. Therefore, future measurements, which aim to expand the sample of galaxies above the discovery threshold, will be essential to test how this parameter varies across the SFGs and SBGs population and to quantify its impact on the diffuse neutrino flux.

Finally, with current Fermi-LAT data we have obtained that high SFR sources have $F_{\rm cal} \gtrsim 50\%$, which is theoretically expected but further data can challenge this concept leading to even a smaller calorimetric fraction. This is crucial because they mainly drive the diffuse $\gamma$-ray and neutrino fluxes. Hence, future analyses and data aiming at directly probing the degree of calorimetry of these sources are fundamental to further constrain the diffuse emission of SFGs and SBGs.

\appendix
\section{Spectral energy distributions}\label{app:SED}

\begin{figure*}[h!]
    \centering
    \includegraphics[width=0.49\columnwidth]{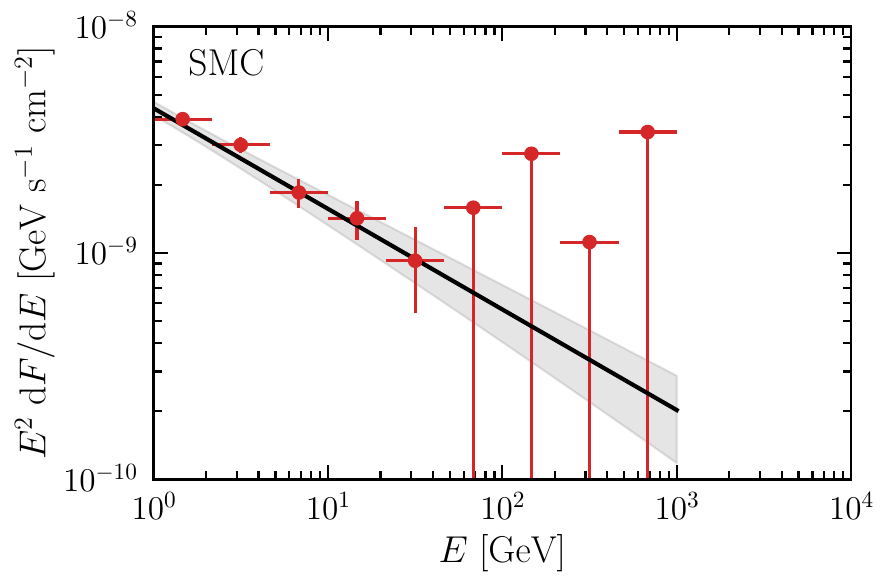}
    \includegraphics[width=0.49\columnwidth]{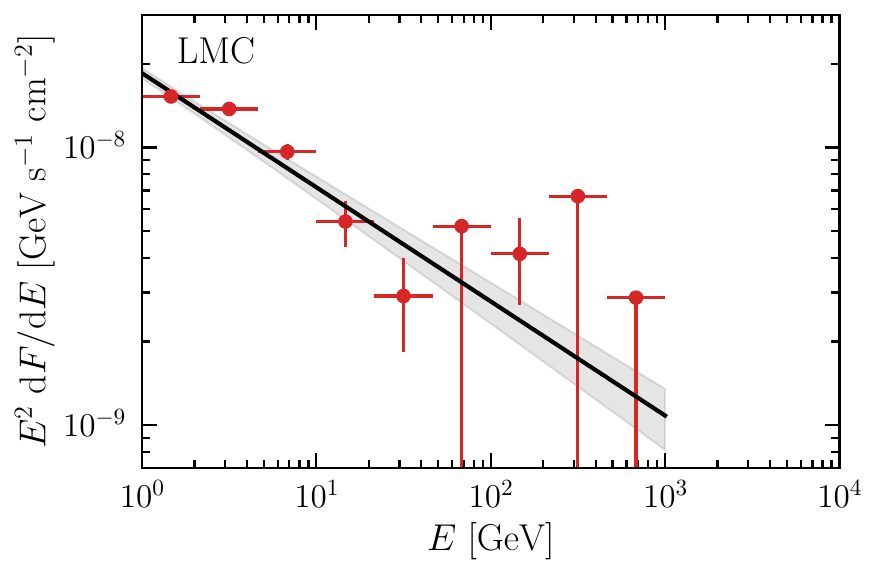}
    \includegraphics[width=0.49\columnwidth]{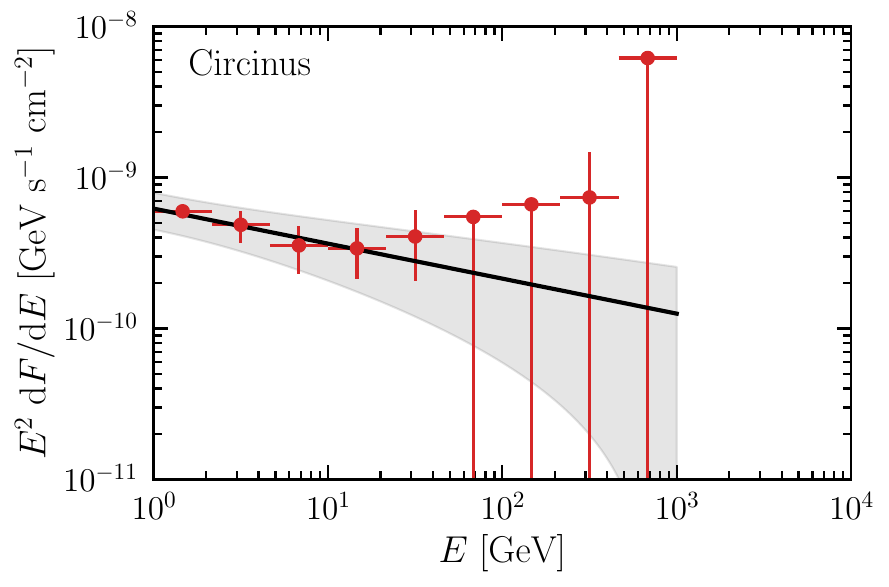}
    \includegraphics[width=0.49\columnwidth]
    {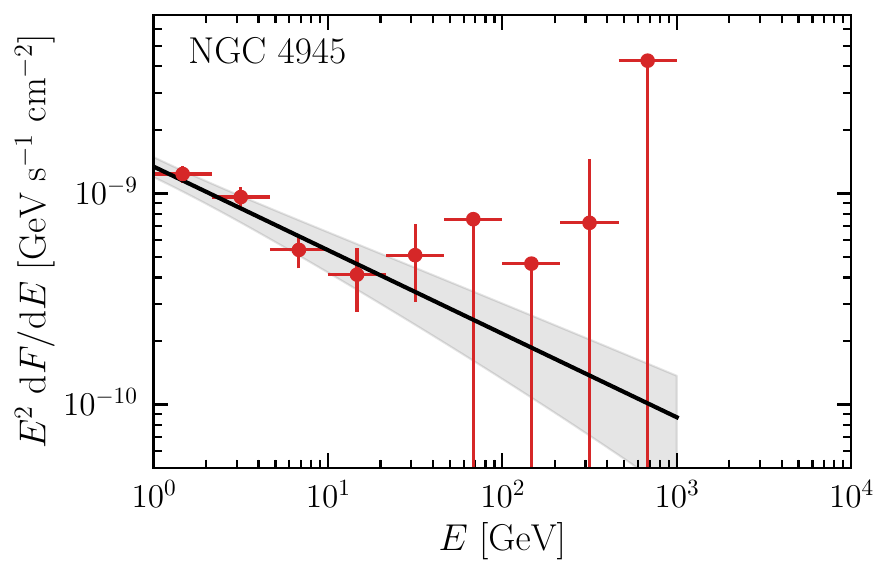}
    \includegraphics[width=0.49\columnwidth]{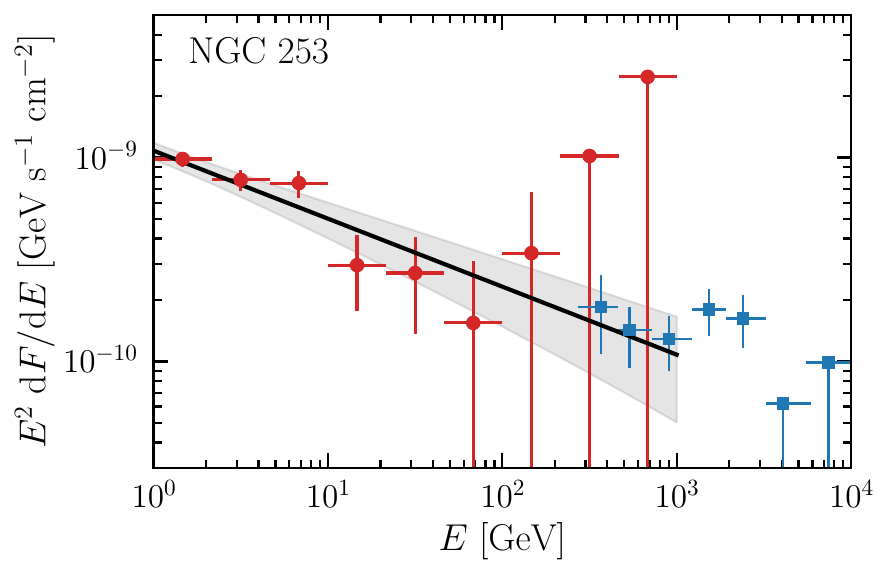}
    \includegraphics[width=0.49\columnwidth]{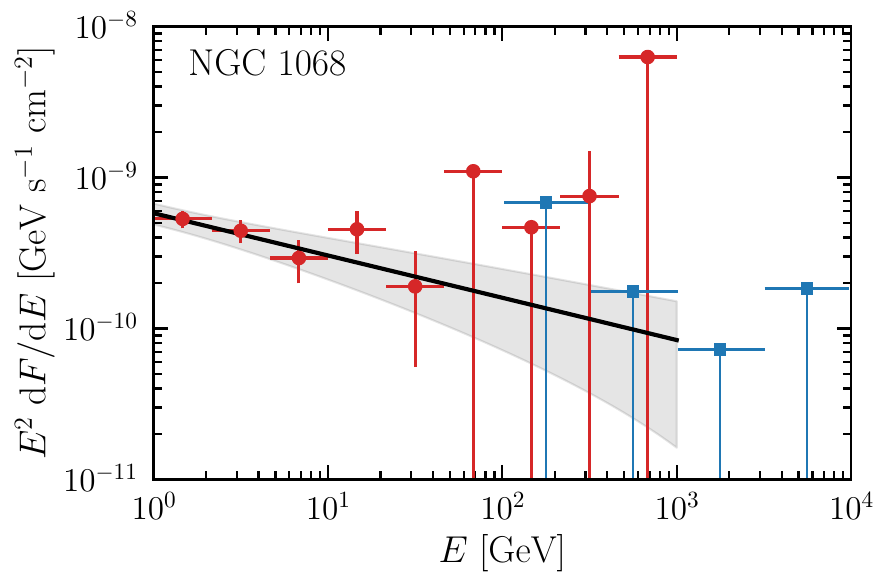}
    \caption{\label{fig:sed_south}
  Collection of the SEDs for sources situated in the southern hemisphere~$(\delta<0^{\circ})$. The red points corresponds to the SED points for each energy bin, while the black line and the grey bands respectively correspond to the best fit and $1\sigma$ band for the fit over the entire energy range. For NGC 253 and NGC 1068, we also report with blue data points the H.E.S.S. measurements~\citep{HESS:2018yqa} and MAGIC upper limits~\citep{MAGIC:2019fvw}, respectively.}
\end{figure*}

\begin{figure*}[h!]
    \centering
    \includegraphics[width=0.49\columnwidth]{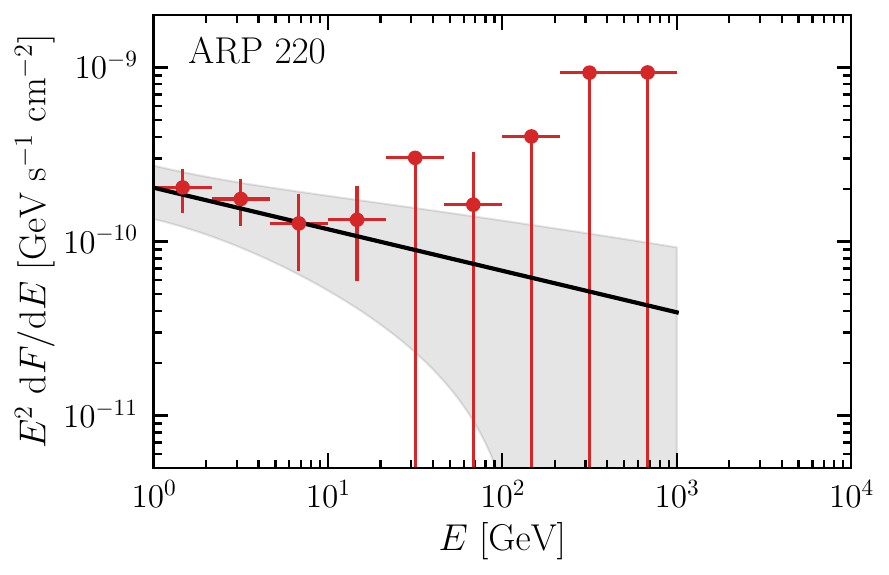}
    \includegraphics[width=0.49\columnwidth]{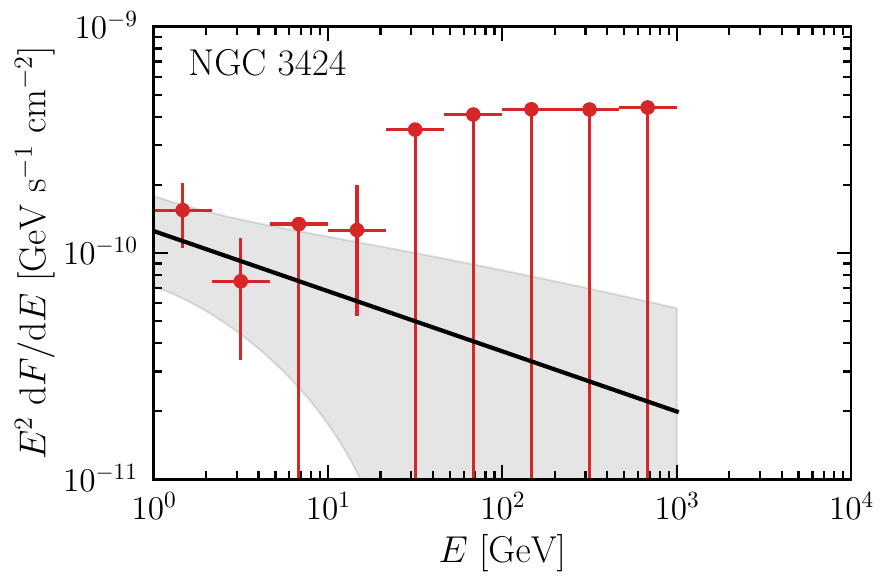}
    \includegraphics[width=0.49\columnwidth]
    {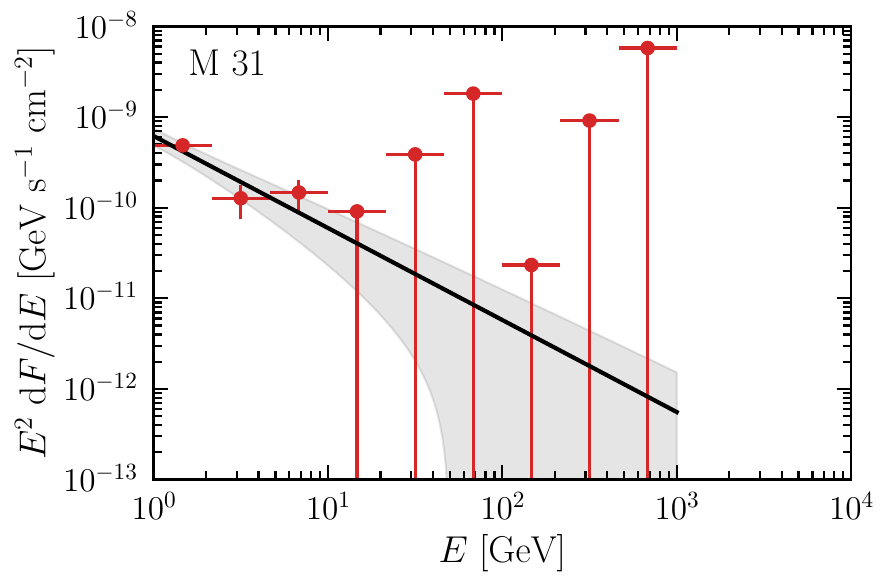}
    \includegraphics[width=0.49\columnwidth]{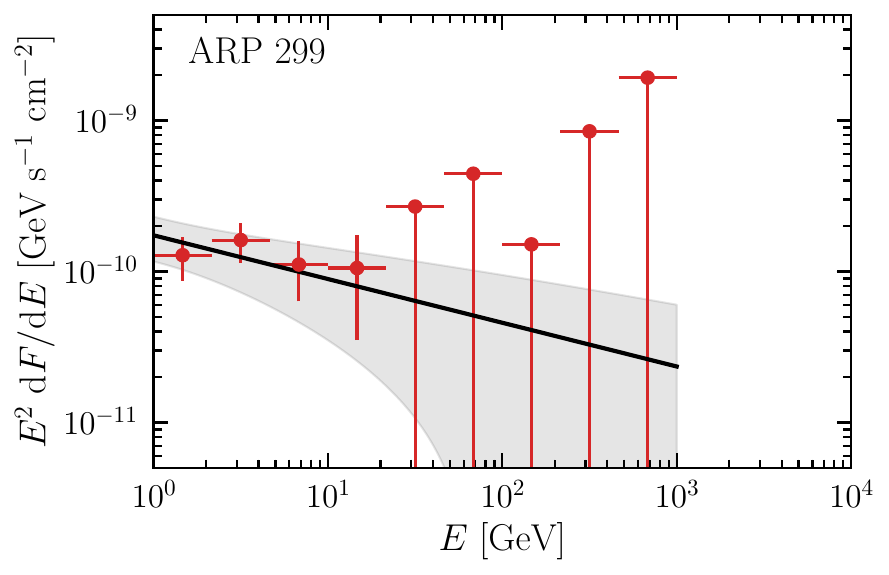}
    \includegraphics[width=0.49\columnwidth]{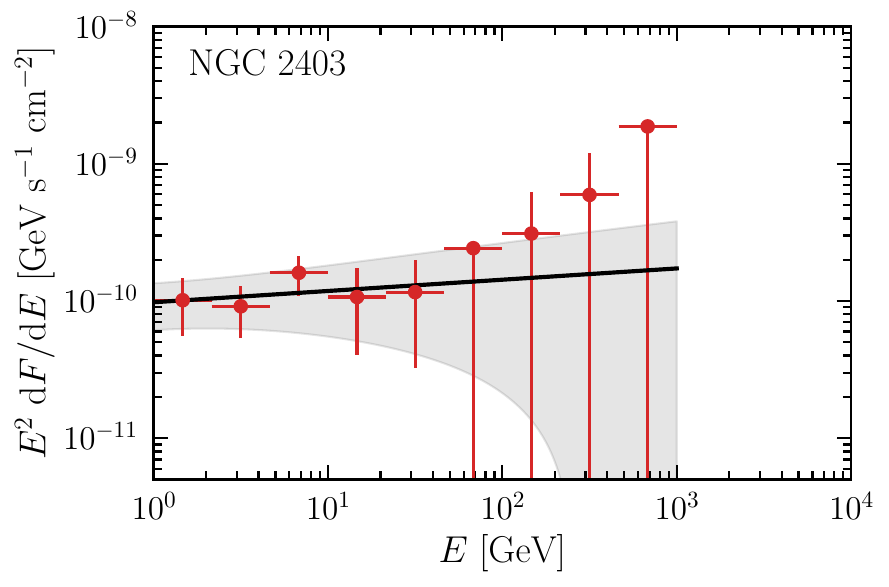}
    \includegraphics[width=0.49\columnwidth]{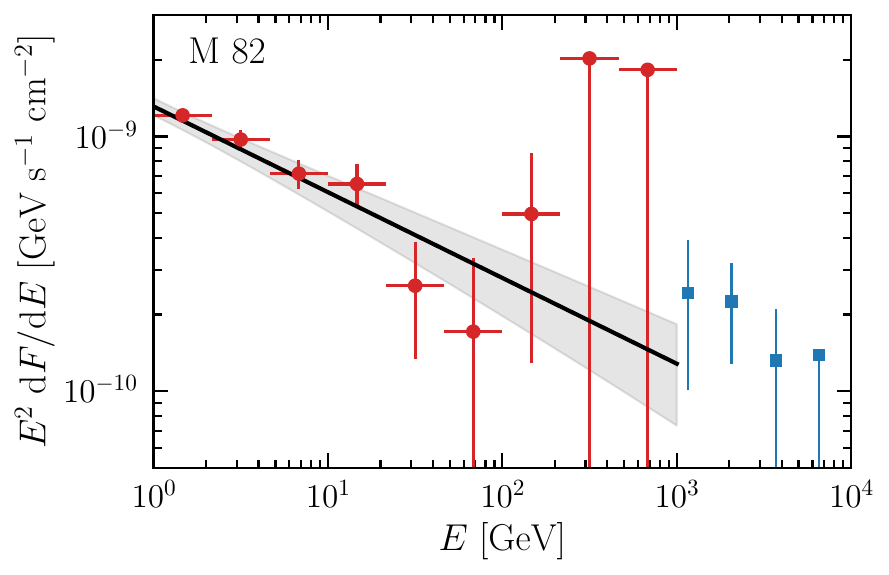}
    \includegraphics[width=0.49\columnwidth]{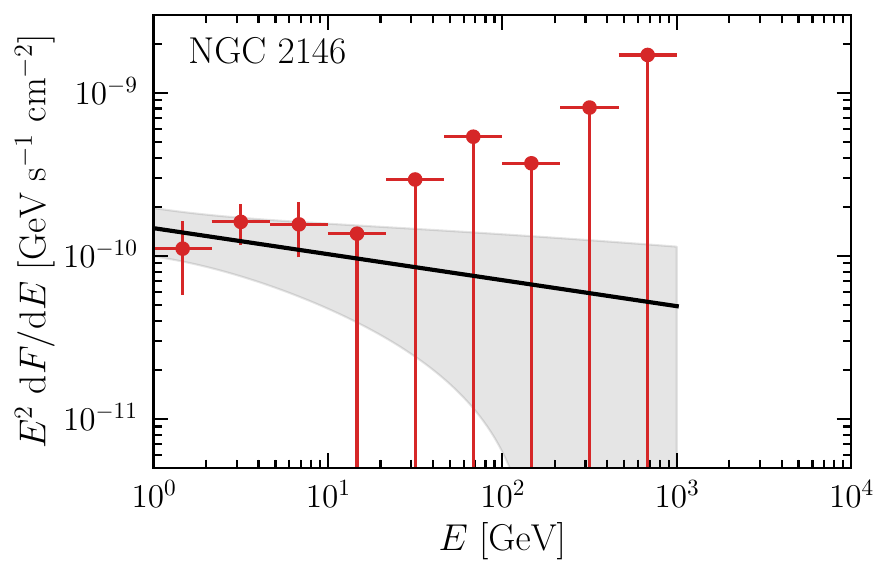}
    \caption{\label{fig:sed_north}
    Collection of the SEDs for sources situated in the northern hemisphere~$(\delta>0^{\circ})$. The red points corresponds to the SED points for each energy bin, while the black line and the grey bands respectively correspond to the best fit and $1\sigma$ band for the fit over the entire energy range. For M82, we also report with blue data points the VERITAS measurements~\citep{2009Natur.462..770V}.}
\end{figure*}

Here we report the spectral energy distributions (SEDs) for the sources above the discovery threshold. We divide the analysed energy range $[1-1000]\, \rm GeV$ in 9 independent bins (3 per decade) and perform a likelihood analysis in each bin fixing the spectral index to $\gamma = 2.0$. If $\rm TS < 4$, then we report the upper limit at $95\%$ CL. Our results are shown in Figs.~\ref{fig:sed_south} and~\ref{fig:sed_north},
where we divide the sources in the northern hemisphere (equatorial declination~$(\delta >0^{\circ}$) and in the southern hemisphere~$(\delta <0^{\circ})$. The red points correspond to the best-fit Fermi-LAT measurements with the $1\sigma$ uncertainty, while the black line and the grey band respectively represent the best-fit and the $1\sigma$ band for the fit over the entire energy range. For M 82, NGC 253 and NGC 1068, we also report the measurements (in blue color) taken by VERITAS~\citep{2009Natur.462..770V}, H.E.S.S.~\citep{HESS:2018yqa} and MAGIC~\citep{MAGIC:2019fvw}, respectively. Finally, for each source (from Tab.~\ref{tab:M82_SED} to Tab.~\ref{tab:LMC_SED}), we report the  obtained TS in each energy bin and if $\rm TS >4$, we report the best-fit value of the SED and its $1\sigma$ uncertainty, otherwise we report its 95\% CL upper limit. 
%
%
\begin{table*}[t!]
    \centering
    \begin{tabular}{c|c|c|c|c}
       Energy range & TS & $E^2~{\rm d}F/{\rm d}E$ & $1\sigma$ error & 95\% CL upper limit  
       \\
       $[\log_{10}(E/{\rm GeV})]$  &  & $[\rm GeV\, \rm cm^{-2}\, \rm s^{-1}]$ & $[\rm GeV\, \rm cm^{-2}\, \rm s^{-1}]$ & $[\rm GeV\, \rm cm^{-2}\, \rm s^{-1}]$  
       \\ \hline
      0.00 - 0.33    & 431 &  $1.21\cdot 10^{-9}$ & $7\cdot 10^{-11}$ & --     \\
      0.33 - 0.66 & 362 &   $9.7\cdot 10^{-10}$& $9\cdot 10^{-11}$   & --  \\
      0.66 - 1.00  & 182 &   $7.2\cdot 10^{-10}$ & $9\cdot 10^{-11}$  & --  \\
      1.00 - 1.33   &   106 &  $6.5\cdot 10^{-10}$  & $1.3\cdot 10^{-10}$ & --    \\
      1.33 - 1.66 &   20 & $2.6\cdot 10^{-10}$ & $1.6\cdot 10^{-10}$ & --    \\
      1.66 - 2.00 & 5  &  $1.7\cdot 10^{-10}$ & $1.6\cdot 10^{-10}$ & --      \\
      2.00 - 2.33 & 11  &   $5\cdot 10^{-10}$ & $3\cdot 10^{-10}$ & --       \\
      2.33 - 2.66 & 3 &   -- & -- &  $2.0\cdot 10^{-9}$    \\
      2.66 - 3.00 &  0 &  -- & -- &  $1.8\cdot 10^{-9}$
    \end{tabular}
    \caption{\textbf{M 82}: the energy range corresponding to each bin, the TS obtained for the source, the best-fit value of the SED and its $1\sigma$ error and finally the the $95\%$ CL upper limit. If $\rm TS >4$, we report the SED best-fit value and its $1\sigma$ error; otherwise we report the $95\%$ CL upper limit.}
    \label{tab:M82_SED}
\end{table*}
\begin{table*}[t!]
    \centering
    \begin{tabular}{c|c|c|c|c}
       Energy range & TS & $E^2~{\rm d}F/{\rm d}E$ & $1\sigma$ error & 95\% CL upper limit  
       \\
       $[\log_{10}(E/{\rm GeV})]$  &  & $[\rm GeV\, \rm cm^{-2}\, \rm s^{-1}]$ & $[\rm GeV\, \rm cm^{-2}\, \rm s^{-1}]$ & $[\rm GeV\, \rm cm^{-2}\, \rm s^{-1}]$  
       \\ \hline
      0.00 - 0.33    &  311 &  $9.8\cdot 10^{-10}$ & $8\cdot 10^{-11}$  & --     \\
      0.33 - 0.66 & 212&$7.7\cdot 10^{-10}$   &  $9\cdot 10^{-11}$ & --  \\
      0.66 - 1.00  & 166 &  $7.5\cdot 10^{-10}$  & $1.1\cdot 10^{-10}$  & --  \\
      1.00 - 1.33   &   24 &  $3.0\cdot 10^{-10}$ & $1.2\cdot 10^{-10}$ & --    \\
      1.33 - 1.66 &  10  &  $2.7\cdot 10^{-10}$ &  $1.3\cdot 10^{-10}$ & --    \\
      1.66 - 2.00 &  5 & $1.5\cdot 10^{-10}$  & $1.5\cdot 10^{-10}$  & --      \\
      2.00 - 2.33 &  7 &  $3.4\cdot 10^{-10}$  & $3.4\cdot 10^{-10}$  & --       \\
      2.33 - 2.66 & 0 &   -- & -- &  $1.0\cdot 10^{-9}$    \\
      2.66 - 3.00 &  0 &  -- & -- &  $2.5\cdot 10^{-9}$
    \end{tabular}
    \caption{\textbf{NGC 253}: the energy range corresponding to each bin, the TS obtained for the source, the best-fit value of the SED and its $1\sigma$ error and finally the the $95\%$ CL upper limit. If $\rm TS >4$, we report the SED best-fit value and its $1\sigma$ error; otherwise we report the $95\%$ CL upper limit.}
    \label{tab:NGC253_SED}
\end{table*}
\begin{table*}[t!]
    \centering
    \begin{tabular}{c|c|c|c|c}
       Energy range & TS & $E^2~{\rm d}F/{\rm d}E$ & $1\sigma$ error & 95\% CL upper limit  
       \\
       $[\log_{10}(E/{\rm GeV})]$  &  & $[\rm GeV\, \rm cm^{-2}\, \rm s^{-1}]$ & $[\rm GeV\, \rm cm^{-2}\, \rm s^{-1}]$ & $[\rm GeV\, \rm cm^{-2}\, \rm s^{-1}]$  
       \\ \hline
      0.00 - 0.33    & 11   & $2.0\cdot 10^{-10}$  & $6\cdot 10^{-11}$  & --     \\
      0.33 - 0.66 &  14 & $1.8\cdot 10^{-10}$ & $5\cdot 10^{-11}$ & --  \\
      0.66 - 1.00  & 10   & $1.3\cdot 10^{-10}$ & $6\cdot 10^{-11}$ & --  \\
      1.00 - 1.33   & 11 &  $1.3\cdot 10^{-10}$ & $7\cdot 10^{-11}$  & --    \\
      1.33 - 1.66 &  3  &  -- & --    & $3.0\cdot 10^{-10}$   \\
      1.66 - 2.00 & 5 & $1.6\cdot 10^{-10}$ & $1.6\cdot 10^{-10}$   & --      \\
      2.00 - 2.33 & 0 & --  & --  &   $4.0\cdot 10^{-10}$     \\
      2.33 - 2.66 & 0 &   -- & -- &  $9.3\cdot 10^{-10}$  \\
      2.66 - 3.00 & 0  &  -- & -- &  $9.3\cdot 10^{-10}$
    \end{tabular}
    \caption{\textbf{ARP 220}: the energy range corresponding to each bin, the TS obtained for the source, the best-fit value of the SED and its $1\sigma$ error and finally the the $95\%$ CL upper limit. If $\rm TS >4$, we report the SED best-fit value and its $1\sigma$ error; otherwise we report the $95\%$ CL upper limit.}
    \label{tab:ARP220_SED}
\end{table*}
\begin{table*}[t!]
    \centering
    \begin{tabular}{c|c|c|c|c}
       Energy range & TS & $E^2~{\rm d}F/{\rm d}E$ & $1\sigma$ error & 95\% CL upper limit  
       \\
       $[\log_{10}(E/{\rm GeV})]$  &  & $[\rm GeV\, \rm cm^{-2}\, \rm s^{-1}]$ & $[\rm GeV\, \rm cm^{-2}\, \rm s^{-1}]$ & $[\rm GeV\, \rm cm^{-2}\, \rm s^{-1}]$  
       \\ \hline
      0.00 - 0.33    &  75  & $5.3\cdot 10^{-10}$ & $7\cdot 10^{-11}$   & --     \\
      0.33 - 0.66 & 79  & $4.4\cdot 10^{-10}$ & $8\cdot 10^{-11}$  & --  \\
      0.66 - 1.00  & 24   & $2.9\cdot 10^{-10}$ & $9\cdot 10^{-1}$  & --  \\
      1.00 - 1.33   & 44 & $4.5\cdot 10^{-10}$  &$1.4\cdot 10^{-10}$  & --    \\
      1.33 - 1.66 & 9   &  $1.9\cdot 10^{-10}$& $1.3\cdot 10^{-10}$ &  --  \\
      1.66 - 2.00 & 0 & -- & --  & $1.1\cdot 10^{-9}$     \\
      2.00 - 2.33 & 0 & --  & --  &  $4.7\cdot 10^{-8}$    \\
      2.33 - 2.66 & 9 &   $7.5\cdot 10^{-10}$& $7.5\cdot 10^{-10}$ & --   \\
      2.66 - 3.00 & 3  &  -- & -- &  $6.3\cdot 10^{-9}$
    \end{tabular}
    \caption{\textbf{NGC 1068}: the energy range corresponding to each bin, the TS obtained for the source, the best-fit value of the SED and its $1\sigma$ error and finally the the $95\%$ CL upper limit. If $\rm TS >4$, we report the SED best-fit value and its $1\sigma$ error; otherwise we report the $95\%$ CL upper limit.}
    \label{tab:NGC1068_SED}
\end{table*}
\begin{table*}[t!]
    \centering
    \begin{tabular}{c|c|c|c|c}
       Energy range & TS & $E^2~{\rm d}F/{\rm d}E$ & $1\sigma$ error & 95\% CL upper limit  
       \\
       $[\log_{10}(E/{\rm GeV})]$  &  & $[\rm GeV\, \rm cm^{-2}\, \rm s^{-1}]$ & $[\rm GeV\, \rm cm^{-2}\, \rm s^{-1}]$ & $[\rm GeV\, \rm cm^{-2}\, \rm s^{-1}]$  
       \\ \hline
      0.00 - 0.33    &  16  &$6.0\cdot 10^{-10}$  &  $6\cdot 10^{-11}$ & --     \\
      0.33 - 0.66 &  21 & $4.9\cdot 10^{-10}$  & $1.2\cdot 10^{-10}$  & --  \\
      0.66 - 1.00  &  15  &  $3.6\cdot 10^{-10}$& $1.3\cdot 10^{-10}$  & --  \\
      1.00 - 1.33   &  9 & $3.4\cdot 10^{-10}$  & $1.3\cdot 10^{-10}$  & --    \\
      1.33 - 1.66 & 10   & $4.1\cdot 10^{-10}$ & $2.0\cdot 10^{-10}$ &  --  \\
      1.66 - 2.00 & 1 &--  & --  &  $5.5\cdot 10^{-10}$    \\
      2.00 - 2.33 & 0 & --  & --  &   $6.6\cdot 10^{-10}$  \\
      2.33 - 2.66 & 5 & $7.4\cdot 10^{-10}$  & $7.4\cdot 10^{-10}$   & --   \\
      2.66 - 3.00 & 3.6  &  -- & -- &  $6.2\cdot 10^{-9}$
    \end{tabular}
    \caption{\textbf{Circinus}: the energy range corresponding to each bin, the TS obtained for the source, the best-fit value of the SED and its $1\sigma$ error and finally the the $95\%$ CL upper limit. If $\rm TS >4$, we report the SED best-fit value and its $1\sigma$ error; otherwise we report the $95\%$ CL upper limit.}
    \label{tab:Circinus_SED}
\end{table*}
\begin{table*}[t!]
    \centering
    \begin{tabular}{c|c|c|c|c}
       Energy range & TS & $E^2~{\rm d}F/{\rm d}E$ & $1\sigma$ error & 95\% CL upper limit  
       \\
       $[\log_{10}(E/{\rm GeV})]$  &  & $[\rm GeV\, \rm cm^{-2}\, \rm s^{-1}]$ & $[\rm GeV\, \rm cm^{-2}\, \rm s^{-1}]$ & $[\rm GeV\, \rm cm^{-2}\, \rm s^{-1}]$  
       \\ \hline
      0.00 - 0.33    & 481   & $3.90\cdot 10^{-9}$& $1.1\cdot 10^{-10}$   & --     \\
      0.33 - 0.66 & 241 & $3.00\cdot 10^{-9}$ & $2.4\cdot 10^{-10}$   & --  \\
      0.66 - 1.00  &  73  &  $1.85\cdot 10^{-9}$& $2.7\cdot 10^{-10}$   & --  \\
      1.00 - 1.33   & 27 & $1.42\cdot 10^{-9}$  & $2.8\cdot 10^{-10}$  & --    \\
      1.33 - 1.66 & 7  & $9\cdot 10^{-10}$  & $3\cdot 10^{-10}$  &  --  \\
      1.66 - 2.00 & 2 &--  & --  &  $1.6\cdot 10^{-9}$   \\
      2.00 - 2.33 & 0 & --  & --  & $2.7\cdot 10^{-9}$ \\
      2.33 - 2.66 & 0 & -- & --  & $1.1\cdot 10^{-9}$  \\
      2.66 - 3.00 & 0  &  -- & -- &  $3.4\cdot 10^{-9}$
    \end{tabular}
    \caption{\textbf{SMC}: the energy range corresponding to each bin, the TS obtained for the source, the best-fit value of the SED and its $1\sigma$ error and finally the the $95\%$ CL upper limit. If $\rm TS >4$, we report the SED best-fit value and its $1\sigma$ error; otherwise we report the $95\%$ CL upper limit.}
    \label{tab:SMC_SED}
\end{table*}
\begin{table*}[t!]
    \centering
    \begin{tabular}{c|c|c|c|c}
       Energy range & TS & $E^2~{\rm d}F/{\rm d}E$ & $1\sigma$ error & 95\% CL upper limit  
       \\
       $[\log_{10}(E/{\rm GeV})]$  &  & $[\rm GeV\, \rm cm^{-2}\, \rm s^{-1}]$ & $[\rm GeV\, \rm cm^{-2}\, \rm s^{-1}]$ & $[\rm GeV\, \rm cm^{-2}\, \rm s^{-1}]$  
       \\ \hline
      0.00 - 0.33    & 58  & $4.9\cdot 10^{-10}$ & $7\cdot 10^{-11}$   & --     \\
      0.33 - 0.66 & 7 & $1.3\cdot 10^{-10}$ & $5\cdot 10^{-11}$   & --  \\
      0.66 - 1.00  &  11  &$1.5\cdot 10^{-10}$ & $6\cdot 10^{-11}$  & --  \\
      1.00 - 1.33   & 3 & --   & -- & $9.1 \cdot 10^{-11}$    \\
      1.33 - 1.66 &  0 & --  & --  & $3.9\cdot 10^{-10}$  \\
      1.66 - 2.00 & 0 &--  & --  &  $1.8\cdot 10^{-9}$   \\
      2.00 - 2.33 & 0 & --  & --  & $2.3\cdot 10^{-11}$  \\
      2.33 - 2.66 & 0 & -- & --  &  $9.2\cdot 10^{-10}$  \\
      2.66 - 3.00 & 3  &  -- & -- &  $5.8\cdot 10^{-9}$
    \end{tabular}
    \caption{\textbf{M 31}: the energy range corresponding to each bin, the TS obtained for the source, the best-fit value of the SED and its $1\sigma$ error and finally the the $95\%$ CL upper limit. If $\rm TS >4$, we report the SED best-fit value and its $1\sigma$ error; otherwise we report the $95\%$ CL upper limit.}
    \label{tab:M31_SED}
\end{table*}
\begin{table*}[t!]
    \centering
    \begin{tabular}{c|c|c|c|c}
       Energy range & TS & $E^2~{\rm d}F/{\rm d}E$ & $1\sigma$ error & 95\% CL upper limit  
       \\
       $[\log_{10}(E/{\rm GeV})]$  &  & $[\rm GeV\, \rm cm^{-2}\, \rm s^{-1}]$ & $[\rm GeV\, \rm cm^{-2}\, \rm s^{-1}]$ & $[\rm GeV\, \rm cm^{-2}\, \rm s^{-1}]$  
       \\ \hline
      0.00 - 0.33    &  5 &  $1.1\cdot 10^{-10}$& $5\cdot 10^{-11}$   & --     \\
      0.33 - 0.66 & 18  & $1.6\cdot 10^{-10}$ & $5\cdot 10^{-11}$  & --  \\
      0.66 - 1.00  & 17 & $1.6\cdot 10^{-10}$ & $6\cdot 10^{-11}$ & --  \\
      1.00 - 1.33   & 0 & --   & -- & $1.4\cdot 10^{-10}$  \\
      1.33 - 1.66 & 3  & --  & --  &  $2.9 \cdot 10^{-10}$ \\
      1.66 - 2.00 & 3.8  &--  & --  &  $5.4\cdot 10^{-10}$  \\
      2.00 - 2.33 & 0 & --  & --  &  $3.7\cdot 10^{-10}$ \\
      2.33 - 2.66 & 0 & -- & --  &  $8.1\cdot 10^{-10}$ \\
      2.66 - 3.00 & 0 &  -- & -- & $1.7\cdot 10^{-9}$
    \end{tabular}
    \caption{\textbf{NGC 2146}: the energy range corresponding to each bin, the TS obtained for the source, the best-fit value of the SED and its $1\sigma$ error and finally the the $95\%$ CL upper limit. If $\rm TS >4$, we report the SED best-fit value and its $1\sigma$ error; otherwise we report the $95\%$ CL upper limit.}
    \label{tab:NGC2146_SED}
\end{table*}
\begin{table*}[t!]
    \centering
    \begin{tabular}{c|c|c|c|c}
       Energy range & TS & $E^2~{\rm d}F/{\rm d}E$ & $1\sigma$ error & 95\% CL upper limit  
       \\
       $[\log_{10}(E/{\rm GeV})]$  &  & $[\rm GeV\, \rm cm^{-2}\, \rm s^{-1}]$ & $[\rm GeV\, \rm cm^{-2}\, \rm s^{-1}]$ & $[\rm GeV\, \rm cm^{-2}\, \rm s^{-1}]$  
       \\ \hline
      0.00 - 0.33    & 9  & $1.3\cdot 10^{-10}$ & $4\cdot 10^{-11}$  & --     \\
      0.33 - 0.66 & 21  & $1.6\cdot 10^{-10}$ & $5\cdot 10^{-11}$  & --  \\
      0.66 - 1.00  &  11 & $1.1\cdot 10^{-10}$ & $5\cdot 10^{-11}$  & --  \\
      1.00 - 1.33   & 7 & $1.1\cdot 10^{-10}$  & $7\cdot 10^{-11}$ &  --\\
      1.33 - 1.66 &  0  & --  & --  & $2.7\cdot 10^{-10}$  \\
      1.66 - 2.00 & 1  &--  & --  &  $4.4\cdot 10^{-10}$  \\
      2.00 - 2.33 & 0 & --  & --  &  $1.5\cdot 10^{-10}$ \\
      2.33 - 2.66 &  0& -- & --  & $8.5\cdot 10^{-10}$ \\
      2.66 - 3.00 &  0&  -- & -- & $1.9\cdot 10^{-9}$
    \end{tabular}
    \caption{\textbf{ARP 299}: the energy range corresponding to each bin, the TS obtained for the source, the best-fit value of the SED and its $1\sigma$ error and finally the the $95\%$ CL upper limit. If $\rm TS >4$, we report the SED best-fit value and its $1\sigma$ error; otherwise we report the $95\%$ CL upper limit.}
    \label{tab:ARP299_SED}
\end{table*}
\begin{table*}[t!]
    \centering
    \begin{tabular}{c|c|c|c|c}
       Energy range & TS & $E^2~{\rm d}F/{\rm d}E$ & $1\sigma$ error & 95\% CL upper limit  
       \\
       $[\log_{10}(E/{\rm GeV})]$  &  & $[\rm GeV\, \rm cm^{-2}\, \rm s^{-1}]$ & $[\rm GeV\, \rm cm^{-2}\, \rm s^{-1}]$ & $[\rm GeV\, \rm cm^{-2}\, \rm s^{-1}]$  
       \\ \hline
      0.00 - 0.33    &  160 & $1.24\cdot 10^{-9}$ & $1.1\cdot 10^{-10}$ & --     \\
      0.33 - 0.66 & 141  & $9.6\cdot 10^{-10}$ & $1.1\cdot 10^{-10}$ & --  \\
      0.66 - 1.00  &  54 & $5.4 \cdot 10^{-10}$  & $1.0\cdot 10^{-10}$ & --  \\
      1.00 - 1.33   & 31 &   $4.1\cdot 10^{-10}$& $1.4\cdot 10^{-10}$ &  --\\
      1.33 - 1.66 &  28  & $5.1\cdot 10^{-10}$ & $2.0 \cdot 10^{-10}$  & --  \\
      1.66 - 2.00 & 3  &--  & --  &  $7.5\cdot 10^{-10}$  \\
      2.00 - 2.33 & 0 & --  & --  & $4.6 \cdot 10^{-10}$ \\
      2.33 - 2.66 &  7& $5 \cdot 10^{-10}$ & $7 \cdot 10^{-10}$   &  -- \\
      2.66 - 3.00 & 0 &  -- & -- & $4.3 \cdot 10^{-9}$
    \end{tabular}
    \caption{\textbf{NGC 4945}: the energy range corresponding to each bin, the TS obtained for the source, the best-fit value of the SED and its $1\sigma$ error and finally the the $95\%$ CL upper limit. If $\rm TS >4$, we report the SED best-fit value and its $1\sigma$ error; otherwise we report the $95\%$ CL upper limit.}
    \label{tab:NGC4945_SED}
\end{table*}
\begin{table*}[t!]
    \centering
    \begin{tabular}{c|c|c|c|c}
       Energy range & TS & $E^2~{\rm d}F/{\rm d}E$ & $1\sigma$ error & 95\% CL upper limit  
       \\
       $[\log_{10}(E/{\rm GeV})]$  &  & $[\rm GeV\, \rm cm^{-2}\, \rm s^{-1}]$ & $[\rm GeV\, \rm cm^{-2}\, \rm s^{-1}]$ & $[\rm GeV\, \rm cm^{-2}\, \rm s^{-1}]$  
       \\ \hline
      0.00 - 0.33    &6 & $1.0\cdot 10^{-10}$ & $5\cdot 10^{-11}$ & --     \\
      0.33 - 0.66 &  6 & $9\cdot 10^{-11}$ & $3\cdot 10^{-11}$  & --  \\
      0.66 - 1.00  & 18  &$1.6\cdot 10^{-10}$  & $5\cdot 10^{-11}$  & --  \\
      1.00 - 1.33   &  7&  $1.0\cdot 10^{-10}$ &$6\cdot 10^{-11}$ &  --\\
      1.33 - 1.66 & 4  & $1\cdot 10^{10}$  & $8\cdot 10^{-11}$   & --  \\
      1.66 - 2.00 &  0 &--  & --  & $2.4 \cdot 10^{-10}$ \\
      2.00 - 2.33 & 6 & $3\cdot 10^{-10}$  & $3\cdot 10^{-10}$  & --  \\
      2.33 - 2.66 &  7& $6\cdot 10^{-10}$ & $6\cdot 10^{-10}$   &  -- \\
      2.66 - 3.00 & 0 &  -- & -- &  $1.9\cdot 10^{-9}$
    \end{tabular}
    \caption{\textbf{NGC 2403}: the energy range corresponding to each bin, the TS obtained for the source, the best-fit value of the SED and its $1\sigma$ error and finally the the $95\%$ CL upper limit. If $\rm TS >4$, we report the SED best-fit value and its $1\sigma$ error; otherwise we report the $95\%$ CL upper limit.}
    \label{tab:NGC2403_SED}
\end{table*}
\begin{table*}[t!]
    \centering
    \begin{tabular}{c|c|c|c|c}
       Energy range & TS & $E^2~{\rm d}F/{\rm d}E$ & $1\sigma$ error & 95\% CL upper limit  
       \\
       $[\log_{10}(E/{\rm GeV})]$  &  & $[\rm GeV\, \rm cm^{-2}\, \rm s^{-1}]$ & $[\rm GeV\, \rm cm^{-2}\, \rm s^{-1}]$ & $[\rm GeV\, \rm cm^{-2}\, \rm s^{-1}]$  
       \\ \hline
      0.00 - 0.33    & 12 & $1.5\cdot 10^{-10}$ & $5\cdot 10^{-11}$  & --     \\
      0.33 - 0.66 & 5 &$8\cdot 10^{-11}$ & $4\cdot 10^{-11}$  & --  \\
      0.66 - 1.00  & 3  & -- & --  & $1.3\cdot 10^{-10}$  \\
      1.00 - 1.33   & 10 &  $1.2\cdot 10^{-10}$  & $7\cdot 10^{-11}$ &  --\\
      1.33 - 1.66 & 3  & -- & --   &  $3.5\cdot 10^{-10}$\\
      1.66 - 2.00 &  0 &--  & --  &$4.1\cdot 10^{-10}$  \\
      2.00 - 2.33 & 0 & -- & --  & $4.3\cdot 10^{-10}$  \\
      2.33 - 2.66 & 0 &  --&  --   &  $4.3\cdot 10^{-10}$  \\
      2.66 - 3.00 & 0 &  -- & -- & $4.4\cdot 10^{-10}$
    \end{tabular}
    \caption{\textbf{NGC 3424}: the energy range corresponding to each bin, the TS obtained for the source, the best-fit value of the SED and its $1\sigma$ error and finally the the $95\%$ CL upper limit. If $\rm TS >4$, we report the SED best-fit value and its $1\sigma$ error; otherwise we report the $95\%$ CL upper limit.}
    \label{tab:NGC3424_SED}
\end{table*}
\begin{table*}[t!]
    \centering
    \begin{tabular}{c|c|c|c|c}
       Energy range & TS & $E^2~{\rm d}F/{\rm d}E$ & $1\sigma$ error & 95\% CL upper limit  
       \\
       $[\log_{10}(E/{\rm GeV})]$  &  & $[\rm GeV\, \rm cm^{-2}\, \rm s^{-1}]$ & $[\rm GeV\, \rm cm^{-2}\, \rm s^{-1}]$ & $[\rm GeV\, \rm cm^{-2}\, \rm s^{-1}]$  
       \\ \hline
      0.00 - 0.33    & 711 & $1.53\cdot 10^{-8}$ &$6\cdot 10^{-10}$  & --     \\
      0.33 - 0.66 & 489 & $1.38\cdot 10^{-8}$ & $6\cdot 10^{-10}$  & --  \\
      0.66 - 1.00  & 173 & $9.7\cdot 10^{-9}$ & $7\cdot 10^{-10}$  & -- \\
      1.00 - 1.33   & 38 & $5.4\cdot 10^{-9}$ & $1.1\cdot 10^{-9}$ &  --\\
      1.33 - 1.66 & 7 & $2.9\cdot 10^{-9}$ & $1.1\cdot 10^{-9}$   &--  \\
      1.66 - 2.00 & 3.7  &--  & --  &  $5.2\cdot 10^{-9}$\\
      2.00 - 2.33 & 21 & $4.1\cdot 10^{-9}$ & $1.4\cdot 10^{-9}$  & --   \\
      2.33 - 2.66 & 0 &  --&  --   & $6.7\cdot 10^{-9}$  \\
      2.66 - 3.00 & 1 &  -- & -- &  $2.9\cdot 10^{-9}$
    \end{tabular}
    \caption{\textbf{LMC}: the energy range corresponding to each bin, the TS obtained for the source, the the best-fit value of the SED and its $1\sigma$ error and finally the the $95\%$ CL upper limit. If $\rm TS >4$, we report the SED best-fit value and its $1\sigma$ error; otherwise we report the $95\%$ CL upper limit.}
    \label{tab:LMC_SED}
\end{table*}

\section{Redshift distribution and the role of ULIRGs}\label{app:diffuse_flux}

Here, we assess which are the redshift and star formation rate values corresponding to the largest contribution to the diffuse $\gamma$-ray and neutrino fluxes. We focus our attention to the diffuse neutrino flux, because neutrinos are not absorbed by the EBL, therefore they maintain the information of the redshift distribution. Furthermore, we fix  $\gamma = 2.3$ and $E_{\nu} = 1\, \rm TeV$, since the results do not change either in terms of the spectral index or the energy. On this regard,  we notice that even though the flux redshifting impacts the high-energy cut-off leading to different conclusions for energies near the cut-off, the final SED is maximum for $E\lesssim 10\, \rm TeV$, making our approximation reasonable.
\begin{figure*}[h!]
    \centering
    \includegraphics[width=0.49\columnwidth]{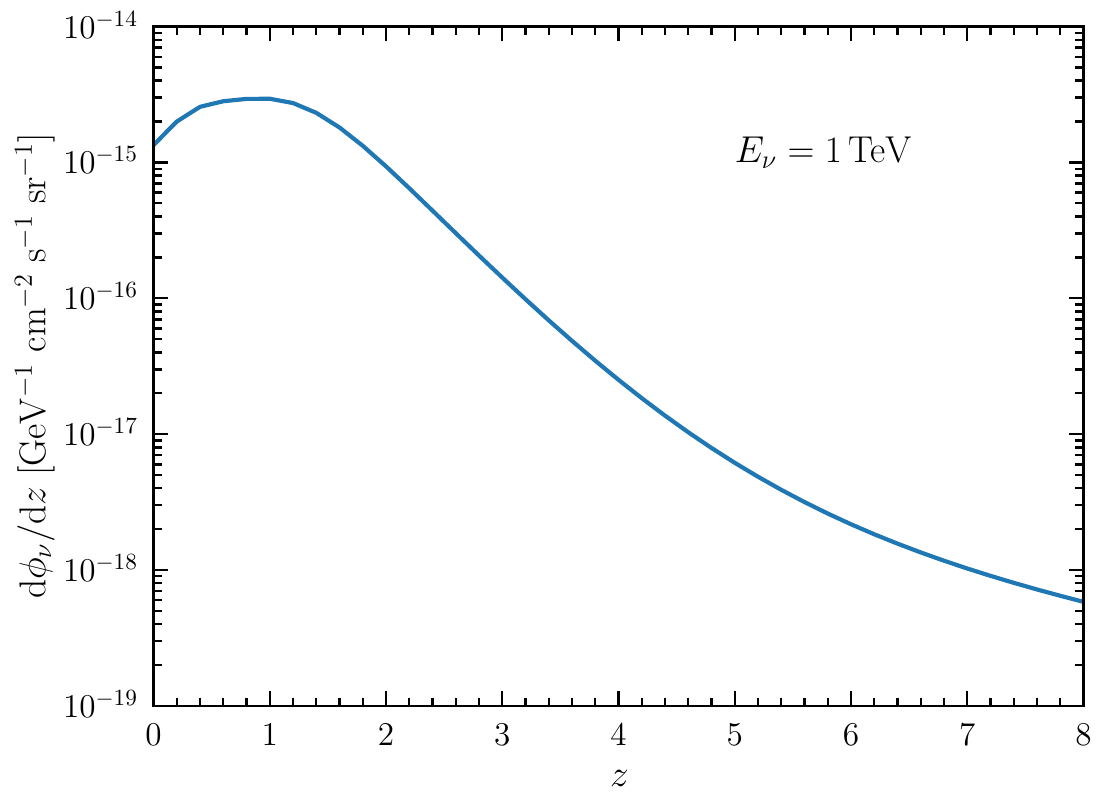}
    \includegraphics[width=0.49\columnwidth]{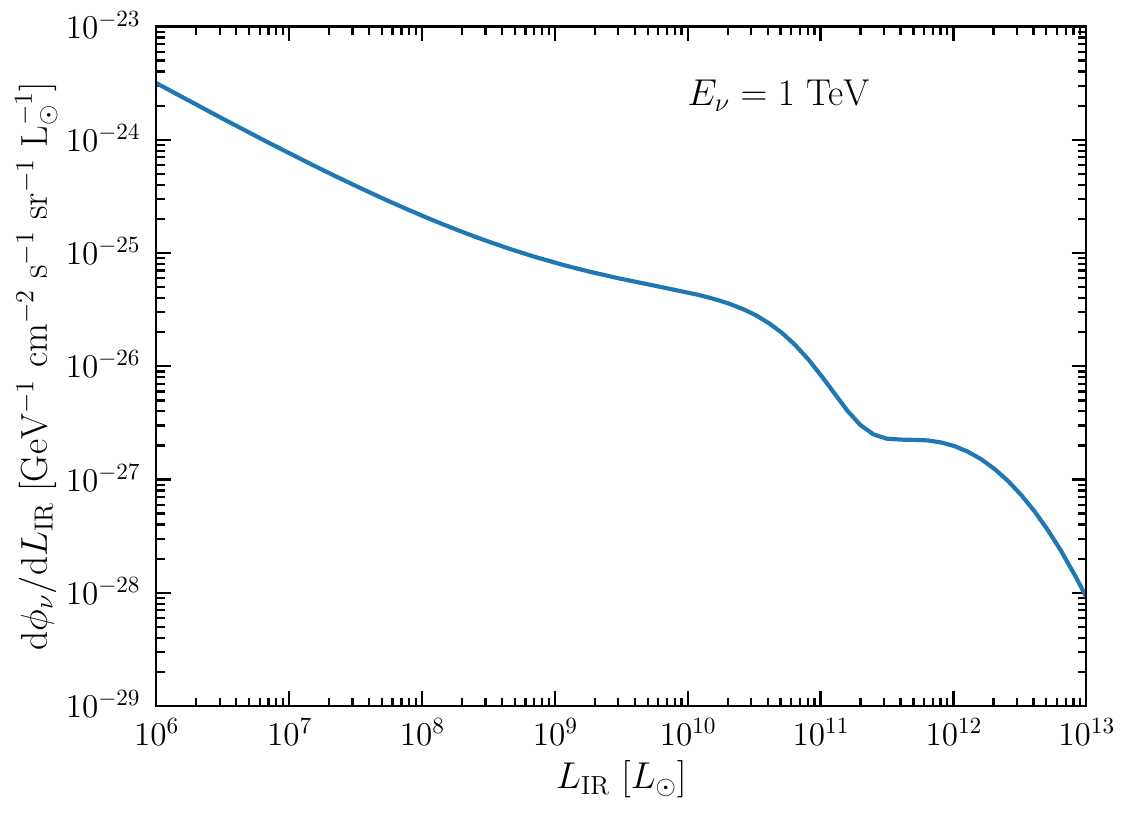}
    \caption{\label{fig:dependence_diffuse_z_L}
    Dependence of the differential diffuse neutrino flux over redshift (left panel) and over the infrared luminosity (right panel) for $E_\nu = 1\, \rm TeV$ and $\gamma = 2.3$.}
\end{figure*}

In the left panel of Fig.~\ref{fig:dependence_diffuse_z_L} we show the redshift distribution of the differential flux, once integrated over the luminosity. It represents the neutrino flux coming at different redshift from the sources of all luminosities. The maximum of the distribution stands for $z \simeq 1$, which represents the maximum of the cosmic star formation rate distribution~\citep{Kim:2023xxh}. In the left panel of Fig.~\ref{fig:dependence_diffuse_z_L} we show the dependence of the differential flux over $L_{\rm IR}$, integrating over all the redshifts. Hence, it quantifies the contribution from all the sources having a given IR luminosity. We stress that even though the maximum of the differential flux stands for the lowest values of $L_{\rm IR}$, the ULIRGs are the ones which contribute most to the total flux. Indeed, the integration over sources with ${\rm SFR} > 100\, M_{\odot}\, {\rm yr^{-1}} (L_{\rm IR} > 7.2\cdot 10^{11}\, L_{\odot})$ provides about 51\% of the total spectrum. Sources with $1 \, M_{\odot} \, {\rm yr^{-1}} < {\rm SFR} < 100\, M_{\odot} \, \rm yr^{-1}$ contribute for 44\% and the remaining 5\% is due to sources with lower star formation rates. Therefore, correctly assessing the calorimetric budget of ULIRGs is fundamental in order to derive correctly the contribution of the entire source population.

\section{Impact of the systematic uncertainty on $R_{SN}$}\label{app:Rsn_systematic_uncertainty}

\begin{figure}[h!]
    \centering
    \includegraphics[width=\columnwidth]{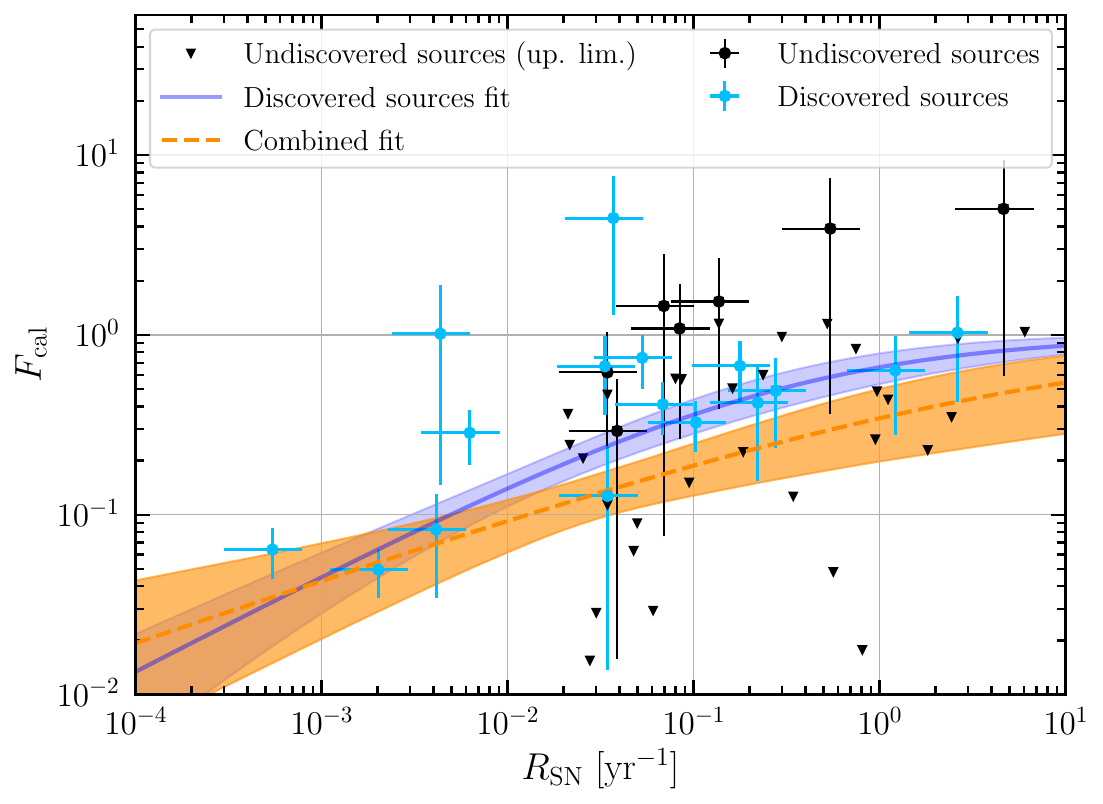}
    \caption{\label{fig:Fcal_vs_rsn_systematic}
   $F_{\rm cal}$ in terms of $R_{\rm SN}$ for the whole sample. Cyan points denote discovered sources, whereas black points denote undiscovered sources. Specifically, for sources exhibiting a flux compatible with zero, we present 95\% CL upper limits indicated by black triangles. For all the sources,  $F_{\rm cal}$  and $R_{\rm SN}$values are reported respectively with 50\% and 45\% systematic uncertainties. We also report the best-fit and the corresponding $1\sigma$ uncertainty band of the fit performed over the whole sample (orange) and over discovered sources (blue).} 
\end{figure}

Here we discuss on the impact of the systematic uncertainty on $\rm R_{\rm SN}$. To this purpose, we assume that the systematic uncertainty on $R_{\rm SN}$ is $\sim 45\%$ instead of 20\% as adopted in the main analysis. Once summed in quadrature with the $10\%$ uncertainty on the distance, it leads to a systematic error of $50\%$ on $F_{\rm cal}$. Using this systematic error, we perform again the fit of the relation in Eq.~\eqref{eq:fcal_fit} and reports the results in Fig.~\ref{fig:Fcal_vs_rsn_systematic}.

For the discovered sources, we find $A = 1.9\pm 1.1 $ and $\beta = 0.54 \pm 0.11$, while for the combined sample $A = 0.5_{-0.2}^{+0.3}$ and $\beta = 0.36\pm 0.11$.
The fits are totally consistent within the $1\sigma$ with the ones presented in the main text. Furthermore, even though larger uncertainties increase the statistical error of the fits, we still find that undiscovered sources are able to constrain $F_{\rm cal}$ reducing its value especially in the range $[0.2-1]\, \rm yr^{-1}$ at $1\sigma$ level. We emphasise that future measurements will be able to reduce the uncertainty on the supernovae explosion rate and they will provide us with a much more constrained correlation function leading to a smaller uncertainty on the diffuse emissions of SFGs and SBGs. 

\section{Power-Law Fit and Comparison with the physically-motivated expression}\label{app:power-law}

\begin{figure}[h!]
    \centering
    \includegraphics[width=\columnwidth]{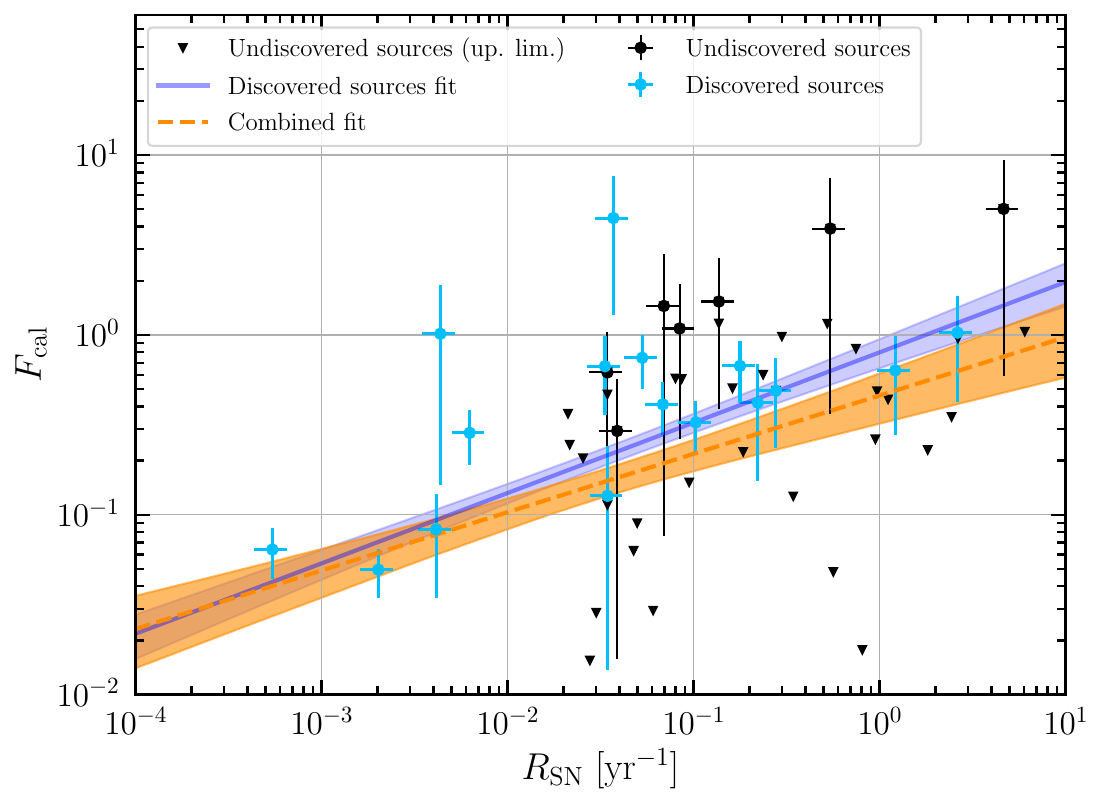}
    \caption{$F_{\rm cal}$ in terms of $R_{\rm SN}$ as in the main text. We report the best-fit and the corresponding $1\sigma$ uncertainty band of the fit performed over the whole sample (orange) and over discovered sources (blue) for the power-law fit. }
    \label{fig:Fcal_vs_Rsn_powerlaw}
\end{figure}

In this section, we discuss the fit using a simple power law expression. Considering only the discovered sources, we obtain $A= 0.8 \pm 0.14$ and $\beta = 0.39\pm 0.04$, while for the whole sample, we obtain $A = 0.45^{+0.11}_{-0.08}$ and $\beta = 0.32\pm 0.05$. Fig. \ref{fig:Fcal_vs_Rsn_powerlaw} shows  the obtained $F_{\rm cal}$ function in terms of $R_{\rm SN}$ for these two fits (best-fit and $1\sigma$  band).

Given the enormous uncertainties on the data, at the moment, there is not any statistical preference for the power-law fit or for Eq.~\ref{eq:fcal_fit}. In particular, we find the following the reduced chi squared:
\begin{itemize}
    \item \textbf{Discovered sources} $\chi^2/\rm d.o.f = 1.58$ (Eq.~\ref{eq:fcal_fit}), $\chi^2/\rm d.o.f = 1.66$ (power-law)
    \item \textbf{Combined sources} $\chi^2/\rm d.o.f = 0.81$ for both Eq.~\ref{eq:fcal_fit} and the power-law case
\end{itemize}
Hence, we conclude that from the data standpoint, it is not possible to exclude a fit from the other, but we advise against the power-law fit, since it is not physical and for high-SFR sources allows for $F_{\rm cal} > 1$.

\section{Impact of Sources Containing AGN contamination and Hint Sources}\label{app:AGN_contamination}

Some of the sources considered in sample B are highly likely contaminated by the AGN activity hosted by the galaxies. These sources are: NGC 2403, NGC 3424 and Circinus Galaxy~\cite{Ajello:2020zna}. Therefore, in this appendix, we discuss how much these sources impact the result of our fits. Furthermore, the three hint sources (M83, M33, NGC 1365) might impact our results in the sense that with more data and a better background description, their detection significance might reduce, in principle changing the results shown in the main text. Furthermore, the M31 galaxy is characterised by a pretty soft spectrum which might not completely be compatible with a CR spectrum dominated by advection. As a result, we remove these 7 sources from the fit and perform the same analysis as shown in the main text. For the discovered sources, we find $A = 2.4 \pm 1.1$ and $\beta = 0.58 \pm 0.08$ with a a reduced chi square at the best-fit value of $\chi^2/\rm d.o.f. = 1.3$. For the combined sample, $A = 0.7^{+0.3}_{-0.2}$ and $\beta = 0.38\pm 0.08$ with $\chi^2/\rm d.o.f. = 0.67$ at the best-fit value. These results are completely consistent with the ones presented in the main text, and consequently, our conclusions remain unchanged.

\acknowledgments

AA is gratefully thankful to Enrico Peretti for fruitful discussion in different stages of the manuscript. 
The authors are supported by the research project TAsP (Theoretical Astroparticle Physics) funded by the Istituto Nazionale di Fisica Nucleare (INFN).




\bibliographystyle{unsrt}

\bibliography{references}

\begin{thebibliography}{10}

\bibitem{Peretti:2018tmo}
Enrico Peretti, Pasquale Blasi, Felix Aharonian, and Giovanni Morlino.
\newblock {Cosmic ray transport and radiative processes in nuclei of starburst
  galaxies}.
\newblock {\em Mon. Not. Roy. Astron. Soc.}, 487(1):168--180, 2019.

\bibitem{Peretti:2019vsj}
Enrico Peretti, Pasquale Blasi, Felix Aharonian, Giovanni Morlino, and Pierre
  Cristofari.
\newblock {Contribution of starburst nuclei to the diffuse gamma-ray and
  neutrino flux}.
\newblock {\em Mon. Not. Roy. Astron. Soc.}, 493(4):5880--5891, 2020.

\bibitem{Ambrosone:2021aaw}
Antonio Ambrosone, Marco Chianese, Damiano F.~G. Fiorillo, Antonio Marinelli,
  and Gennaro Miele.
\newblock {Could Nearby Star-forming Galaxies Light Up the Pointlike Neutrino
  Sky?}
\newblock {\em Astrophys. J. Lett.}, 919(2):L32, 2021.

\bibitem{Ambrosone:2020evo}
Antonio Ambrosone, Marco Chianese, Damiano F.~G. Fiorillo, Antonio Marinelli,
  Gennaro Miele, and Ofelia Pisanti.
\newblock {Starburst galaxies strike back: a multi-messenger analysis with
  Fermi-LAT and IceCube data}.
\newblock {\em Mon. Not. Roy. Astron. Soc.}, 503(3):4032--4049, 2021.

\bibitem{2020A&A...641A.147K}
P.~{Kornecki}, L.~J. {Pellizza}, S.~{del Palacio}, A.~L. {M{\"u}ller}, J.~F.
  {Albacete-Colombo}, and G.~E. {Romero}.
\newblock {{\ensuremath{\gamma}}-ray/infrared luminosity correlation of
  star-forming galaxies}.
\newblock {\em \aap}, 641:A147, September 2020.

\bibitem{Kornecki:2021xiy}
P.~Kornecki, E.~Peretti, S.~del Palacio, P.~Benaglia, and L.~J. Pellizza.
\newblock {Exploring the physics behind the non-thermal emission from
  star-forming galaxies detected in \ensuremath{\gamma} rays}.
\newblock {\em Astron. Astrophys.}, 657:A49, 2022.

\bibitem{Merckx:2023kvn}
Yarno Merckx, Pablo Correa, Krijn~D. de~Vries, Kumiko Kotera, George~C. Privon,
  and Nick van Eijndhoven.
\newblock {Investigating starburst-driven neutrino emission from galaxies in
  the Great Observatories All-Sky LIRG Survey}.
\newblock {\em Phys. Rev. D}, 108(2):023015, 2023.

\bibitem{Werhahn:2021bal}
Maria Werhahn, Christoph Pfrommer, Philipp Girichidis, and Georg Winner.
\newblock {Cosmic rays and non-thermal emission in simulated galaxies
  \textendash{} II. \ensuremath{\gamma}-ray maps, spectra, and the
  far-infrared\textendash{}\ensuremath{\gamma}-ray relation}.
\newblock {\em Mon. Not. Roy. Astron. Soc.}, 505(3):3295--3313, 2021.

\bibitem{Werhahn:2021jvy}
Maria Werhahn, Christoph Pfrommer, and Philipp Girichidis.
\newblock {Cosmic rays and non-thermal emission in simulated galaxies
  \textendash{} III. Probing cosmic-ray calorimetry with radio spectra and the
  FIR\textendash{}radio correlation}.
\newblock {\em Mon. Not. Roy. Astron. Soc.}, 508(3):4072--4095, 2021.

\bibitem{Werhahn:2023osl}
Maria Werhahn, Philipp Girichidis, Christoph Pfrommer, and Joseph Whittingham.
\newblock {Gamma-ray emission from spectrally resolved cosmic rays in
  galaxies}.
\newblock {\em Mon. Not. Roy. Astron. Soc.}, 525(3):4437--4455, 2023.

\bibitem{Nunez-Castineyra:2022rjc}
A.~Nu\~nez Casti\~neyra, I.~A. Grenier, F.~Bournaud, Y.~Dubois, F.~R.
  Kamal~Youssef, and P.~Hennebelle.
\newblock {Cosmic-ray diffusion and the multi-phase interstellar medium in a
  dwarf galaxy. I. Large-scale properties and $\gamma$-ray luminosities}.
\newblock 5 2022.

\bibitem{2012ApJ...755..164A}
M.~{Ackermann}, M.~{Ajello}, A.~{Allafort}, L.~{Baldini}, J.~{Ballet},
  D.~{Bastieri}, K.~{Bechtol}, R.~{Bellazzini}, B.~{Berenji}, E.~D. {Bloom},
  E.~{Bonamente}, A.~W. {Borgland}, A.~{Bouvier}, J.~{Bregeon}, M.~{Brigida},
  P.~{Bruel}, R.~{Buehler}, S.~{Buson}, G.~A. {Caliandro}, R.~A. {Cameron},
  P.~A. {Caraveo}, J.~M. {Casandjian}, C.~{Cecchi}, E.~{Charles},
  A.~{Chekhtman}, C.~C. {Cheung}, J.~{Chiang}, A.~N. {Cillis}, S.~{Ciprini},
  R.~{Claus}, J.~{Cohen-Tanugi}, J.~{Conrad}, S.~{Cutini}, F.~{de Palma}, C.~D.
  {Dermer}, S.~W. {Digel}, E.~do Couto~e. {Silva}, P.~S. {Drell},
  A.~{Drlica-Wagner}, C.~{Favuzzi}, S.~J. {Fegan}, P.~{Fortin}, Y.~{Fukazawa},
  S.~{Funk}, P.~{Fusco}, F.~{Gargano}, D.~{Gasparrini}, S.~{Germani},
  N.~{Giglietto}, F.~{Giordano}, T.~{Glanzman}, G.~{Godfrey}, I.~A. {Grenier},
  S.~{Guiriec}, M.~{Gustafsson}, D.~{Hadasch}, M.~{Hayashida}, E.~{Hays}, R.~E.
  {Hughes}, G.~{J{\'o}hannesson}, A.~S. {Johnson}, T.~{Kamae}, H.~{Katagiri},
  J.~{Kataoka}, J.~{Kn{\"o}dlseder}, M.~{Kuss}, J.~{Lande}, F.~{Longo},
  F.~{Loparco}, B.~{Lott}, M.~N. {Lovellette}, P.~{Lubrano}, G.~M. {Madejski},
  P.~{Martin}, M.~N. {Mazziotta}, J.~E. {McEnery}, P.~F. {Michelson},
  T.~{Mizuno}, C.~{Monte}, M.~E. {Monzani}, A.~{Morselli}, I.~V. {Moskalenko},
  S.~{Murgia}, S.~{Nishino}, J.~P. {Norris}, E.~{Nuss}, M.~{Ohno}, T.~{Ohsugi},
  A.~{Okumura}, N.~{Omodei}, E.~{Orlando}, M.~{Ozaki}, D.~{Parent},
  M.~{Persic}, M.~{Pesce-Rollins}, V.~{Petrosian}, M.~{Pierbattista},
  F.~{Piron}, G.~{Pivato}, T.~A. {Porter}, S.~{Rain{\`o}}, R.~{Rando},
  M.~{Razzano}, A.~{Reimer}, O.~{Reimer}, S.~{Ritz}, M.~{Roth}, C.~{Sbarra},
  C.~{Sgr{\`o}}, E.~J. {Siskind}, G.~{Spandre}, P.~{Spinelli}, {\L}ukasz
  {Stawarz}, A.~W. {Strong}, H.~{Takahashi}, T.~{Tanaka}, J.~B. {Thayer},
  L.~{Tibaldo}, M.~{Tinivella}, D.~F. {Torres}, G.~{Tosti}, E.~{Troja},
  Y.~{Uchiyama}, J.~{Vandenbroucke}, G.~{Vianello}, V.~{Vitale}, A.~P. {Waite},
  M.~{Wood}, and Z.~{Yang}.
\newblock {GeV Observations of Star-forming Galaxies with the Fermi Large Area
  Telescope}.
\newblock {\em \apj}, 755(2):164, August 2012.

\bibitem{Rojas-Bravo:2016val}
Cesar Rojas-Bravo and Miguel Araya.
\newblock {Search for gamma-ray emission from star-forming galaxies with Fermi
  LAT}.
\newblock {\em Mon. Not. Roy. Astron. Soc.}, 463(1):1068--1073, 2016.

\bibitem{Ajello:2020zna}
M.~Ajello, M.~Di~Mauro, V.~S. Paliya, and S.~Garrappa.
\newblock {The $\gamma$-Ray Emission of Star-forming Galaxies}.
\newblock {\em Astrophys. J.}, 894(2):88, 2020.

\bibitem{Xiang:2023dww}
Yunchuan Xiang, Qingquan Jiang, and Xiaofei Lan.
\newblock {Fermi-LAT Detection of a New Starburst Galaxy Candidate: IRAS
  13052-5711}.
\newblock {\em Astrophys. J.}, 953(1):95, 2023.

\bibitem{Lacki:2013ry}
Brian~C. Lacki and Rainer Beck.
\newblock {The Equipartition Magnetic Field Formula in Starburst Galaxies:
  Accounting for Pionic Secondaries and Strong Energy Losses}.
\newblock {\em Mon. Not. Roy. Astron. Soc.}, 430:3171, 2013.

\bibitem{Krumholz:2019uom}
Mark~R. Krumholz, Roland~M. Crocker, Siyao Xu, A.~Lazarian, M.~T. Rosevear, and
  Jasper Bedwell-Wilson.
\newblock {Cosmic ray transport in starburst galaxies}.
\newblock {\em Mon. Not. Roy. Astron. Soc.}, 493(2):2817--2833, 2020.

\bibitem{Roth:2021lvk}
Matt~A. Roth, Mark~R. Krumholz, Roland~M. Crocker, and Silvia Celli.
\newblock {The diffuse \ensuremath{\gamma}-ray background is dominated by
  star-forming galaxies}.
\newblock {\em Nature}, 597(7876):341--344, 2021.

\bibitem{Roth:2022hxc}
Matt~A. Roth, Mark~R. Krumholz, Roland~M. Crocker, and Todd~A. Thompson.
\newblock {congruents (COsmic ray, Neutrino, Gamma-ray, and Radio Non-Thermal
  Spectra) \textendash{} I. A predictive model for galactic non-thermal
  emission}.
\newblock {\em Mon. Not. Roy. Astron. Soc.}, 523(2):2608--2629, 2023.

\bibitem{Ambrosone:2022fip}
Antonio Ambrosone, Marco Chianese, Damiano F.~G. Fiorillo, Antonio Marinelli,
  and Gennaro Miele.
\newblock {Observable signatures of cosmic rays transport in Starburst Galaxies
  on gamma-ray and neutrino observations}.
\newblock {\em Mon. Not. Roy. Astron. Soc.}, 515(4):5389--5399, 2022.

\bibitem{Pfrommer:2017jau}
C.~Pfrommer, R.~Pakmor, C.~M. Simpson, and V.~Springel.
\newblock {Simulating Gamma-ray Emission in Star-forming Galaxies}.
\newblock {\em Astrophys. J. Lett.}, 847(2):L13, 2017.

\bibitem{Crocker:2020yub}
Roland~M. Crocker, Mark~R. Krumholz, and Todd~A. Thompson.
\newblock {Cosmic rays across the star-forming galaxy sequence \textendash{} I.
  Cosmic ray pressures and calorimetry}.
\newblock {\em Mon. Not. Roy. Astron. Soc.}, 502(1):1312--1333, 2021.

\bibitem{Kim:2023xxh}
Seong~Jin Kim, Tomotsugu Goto, Chih-Teng Ling, Cossas K.~W. Wu, Tetsuya
  Hashimoto, Ece Kilerci, Simon C.~C. Ho, Yuri Uno, Po-Ya Wang, and Yu-Wei Lin.
\newblock {Cosmic star-formation history and black hole accretion history
  inferred from the JWST mid-infrared source counts}.
\newblock 12 2023.

\bibitem{Fermi-LAT:2014ryh}
M.~Ackermann et~al.
\newblock {The spectrum of isotropic diffuse gamma-ray emission between 100 MeV
  and 820 GeV}.
\newblock {\em Astrophys. J.}, 799:86, 2015.

\bibitem{IceCube:2020acn}
M.~G. Aartsen et~al.
\newblock {Characteristics of the diffuse astrophysical electron and tau
  neutrino flux with six years of IceCube high energy cascade data}.
\newblock {\em Phys. Rev. Lett.}, 125(12):121104, 2020.

\bibitem{Inoue:2019fil}
Yoshiyuki Inoue, Dmitry Khangulyan, Susumu Inoue, and Akihiro Doi.
\newblock {On high-energy particles in accretion disk coronae of supermassive
  black holes: implications for MeV gamma rays and high-energy neutrinos from
  AGN cores}.
\newblock 4 2019.

\bibitem{Murase:2023ccp}
Kohta Murase, Christopher~M. Karwin, Shigeo~S. Kimura, Marco Ajello, and Sara
  Buson.
\newblock {Sub-GeV Gamma Rays from Nearby Seyfert Galaxies and Implications for
  Coronal Neutrino Emission}.
\newblock {\em Astrophys. J. Lett.}, 961(2):L34, 2024.

\bibitem{Kheirandish:2021wkm}
Ali Kheirandish, Kohta Murase, and Shigeo~S. Kimura.
\newblock {High-energy Neutrinos from Magnetized Coronae of Active Galactic
  Nuclei and Prospects for Identification of Seyfert Galaxies and Quasars in
  Neutrino Telescopes}.
\newblock {\em Astrophys. J.}, 922(1):45, 2021.

\bibitem{Fermi-LAT:2019yla}
S.~Abdollahi et~al.
\newblock {$Fermi$ Large Area Telescope Fourth Source Catalog}.
\newblock {\em Astrophys. J. Suppl.}, 247(1):33, 2020.

\bibitem{Malyshev:2023xya}
Denys Malyshev and Lars Mohrmann.
\newblock {Analysis Methods for Gamma-ray Astronomy}.
\newblock 9 2023.

\bibitem{Fermi-LAT:2022byn}
Soheila Abdollahi et~al.
\newblock {Incremental Fermi Large Area Telescope Fourth Source Catalog}.
\newblock {\em Astrophys. J. Supp.}, 260(2):53, 2022.

\bibitem{Ballet:2023qzs}
J.~Ballet, P.~Bruel, T.~H. Burnett, and B.~Lott.
\newblock {Fermi Large Area Telescope Fourth Source Catalog Data Release 4
  (4FGL-DR4)}.
\newblock 7 2023.

\bibitem{Blanco:2023dfp}
Carlos Blanco, Dan Hooper, Tim Linden, and Elena Pinetti.
\newblock {On the Neutrino and Gamma-Ray Emission from NGC 1068}.
\newblock 7 2023.

\bibitem{TelescopeArray:2023sbd}
R.~U. Abbasi et~al.
\newblock {An extremely energetic cosmic ray observed by a surface detector
  array}.
\newblock {\em Science}, 382:903--907, 2023.

\bibitem{Fermi-LAT:2017ztt}
M.~Ackermann et~al.
\newblock {Observations of M31 and M33 with the Fermi Large Area Telescope: A
  Galactic Center Excess in Andromeda?}
\newblock {\em Astrophys. J.}, 836(2):208, 2017.

\bibitem{2003ApJ...586..794B}
Eric~F. {Bell}.
\newblock {Estimating Star Formation Rates from Infrared and Radio
  Luminosities: The Origin of the Radio-Infrared Correlation}.
\newblock {\em \apj}, 586(2):794--813, April 2003.

\bibitem{Evoli:2019wwu}
Carmelo Evoli, Roberto Aloisio, and Pasquale Blasi.
\newblock {Galactic cosmic rays after the AMS-02 observations}.
\newblock {\em Phys. Rev. D}, 99(10):103023, 2019.

\bibitem{Caprioli:2020spz}
Damiano Caprioli, Colby~C. Haggerty, and Pasquale Blasi.
\newblock {Kinetic Simulations of Cosmic-Ray-Modified Shocks II: Particle
  Spectra}.
\newblock {\em Astrophys. J.}, 905(1):2, 2020.

\bibitem{Kelner:2006tc}
S.~R. Kelner, Felex~A. Aharonian, and V.~V. Bugayov.
\newblock {Energy spectra of gamma-rays, electrons and neutrinos produced at
  proton-proton interactions in the very high energy regime}.
\newblock {\em Phys. Rev. D}, 74:034018, 2006.
\newblock [Erratum: Phys.Rev.D 79, 039901 (2009)].

\bibitem{Franceschini:2017iwq}
Alberto Franceschini and Giulia Rodighiero.
\newblock {The extragalactic background light revisited and the cosmic
  photon-photon opacity}.
\newblock {\em Astron. Astrophys.}, 603:A34, 2017.

\bibitem{Inoue:2022yak}
Susumu Inoue, Matteo Cerruti, Kohta Murase, and Ruo-Yu Liu.
\newblock {Multimessenger emission from winds and tori in active galactic
  nuclei}.
\newblock {\em PoS}, ICRC2023:1161, 2023.

\bibitem{Liu:2017bjr}
Ruo-Yu Liu, Kohta Murase, Susumu Inoue, Chong Ge, and Xiang-Yu Wang.
\newblock {Can winds driven by active galactic nuclei account for the
  extragalactic gamma-ray and neutrino backgrounds?}
\newblock {\em Astrophys. J.}, 858(1):9, 2018.

\bibitem{Senno:2015tra}
Nicholas Senno, Peter M\'esz\'aros, Kohta Murase, Philipp Baerwald, and
  Martin~J. Rees.
\newblock {Extragalactic star-forming galaxies with hypernovae and supernovae
  as high-energy neutrino and gamma-ray sources: the case of the 10 TeV
  neutrino data}.
\newblock {\em Astrophys. J.}, 806(1):24, 2015.

\bibitem{Peretti:2023xqk}
Enrico Peretti, Alessandra Lamastra, Francesco~Gabriele Saturni, Markus Ahlers,
  Pasquale Blasi, Giovanni Morlino, and Pierre Cristofari.
\newblock {Diffusive shock acceleration at EeV and associated multimessenger
  flux from ultra-fast outflows driven by active galactic nuclei}.
\newblock {\em Mon. Not. Roy. Astron. Soc.}, 526(1):181--192, 2023.

\bibitem{Kennicutt:1998zb}
Robert~C. Kennicutt, Jr.
\newblock {Star formation in galaxies along the Hubble sequence}.
\newblock {\em Ann. Rev. Astron. Astrophys.}, 36:189--231, 1998.

\bibitem{2021ApJ...908...61K}
Jr. {Kennicutt}, Robert~C. and Mithi A.~C. {De Los Reyes}.
\newblock {Revisiting the Integrated Star Formation Law. II. Starbursts and the
  Combined Global Schmidt Law}.
\newblock {\em \apj}, 908(1):61, February 2021.

\bibitem{2018Galax...6..138R}
David S.~N. {Rupke}.
\newblock {A Review of Recent Observations of Galactic Winds Driven by Star
  Formation}.
\newblock {\em Galaxies}, 6(4):138, December 2018.

\bibitem{Marinelli:2018lzs}
Antonio Marinelli, Daniele Gaggero, Dario Grasso, Marco Taoso, Alfredo Urbano,
  and Sofia Ventura.
\newblock {High Energy Neutrino expectations from the Central Molecular Zone}.
\newblock {\em PoS}, ICRC2017:939, 2018.

\bibitem{HESS:2017tce}
H.~Abdalla et~al.
\newblock {Characterising the VHE diffuse emission in the central 200 parsecs
  of our Galaxy with H.E.S.S}.
\newblock {\em Astron. Astrophys.}, 612:A9, 2018.

\bibitem{2023MNRAS.518.6273S}
Yoshiaki {Sofue}.
\newblock {Supernova-remnant origin of the Galactic-Centre filaments}.
\newblock {\em \mnras}, 518(4):6273--6292, February 2023.

\bibitem{2011MNRAS.416...70G}
C.~{Gruppioni}, F.~{Pozzi}, G.~{Zamorani}, and C.~{Vignali}.
\newblock {Modelling galaxy and AGN evolution in the infrared: black hole
  accretion versus star formation activity}.
\newblock {\em \mnras}, 416(1):70--86, September 2011.

\bibitem{IceCube:2020wum}
R.~Abbasi et~al.
\newblock {The IceCube high-energy starting event sample: Description and flux
  characterization with 7.5 years of data}.
\newblock {\em Phys. Rev. D}, 104:022002, 2021.

\bibitem{Zeng:2021mhf}
Houdun Zeng, Yuliang Xin, Shuinai Zhang, and Siming Liu.
\newblock {TeV Cosmic-Ray Nucleus Acceleration in Shell-type Supernova Remnants
  with Hard $\gamma$-Ray Spectra}.
\newblock {\em Astrophys. J.}, 910(1):78, 2021.

\bibitem{Morlino:2017gck}
Giovanni Morlino.
\newblock {Supernova Remnant-Cosmic Ray connection: a modern view}.
\newblock {\em IAU Symp.}, 331:230--241, 2017.

\bibitem{Bechtol:2015uqb}
Keith Bechtol, Markus Ahlers, Mattia Di~Mauro, Marco Ajello, and Justin
  Vandenbroucke.
\newblock {Evidence against star-forming galaxies as the dominant source of
  IceCube neutrinos}.
\newblock {\em Astrophys. J.}, 836(1):47, 2017.

\bibitem{Tamborra:2014xia}
Irene Tamborra, Shin'ichiro Ando, and Kohta Murase.
\newblock {Star-forming galaxies as the origin of diffuse high-energy
  backgrounds: Gamma-ray and neutrino connections, and implications for
  starburst history}.
\newblock {\em JCAP}, 09:043, 2014.

\bibitem{2009Natur.462..770V}
{VERITAS Collaboration}, V.~A. {Acciari}, E.~{Aliu}, T.~{Arlen}, T.~{Aune},
  M.~{Bautista}, M.~{Beilicke}, W.~{Benbow}, D.~{Boltuch}, S.~M. {Bradbury},
  J.~H. {Buckley}, V.~{Bugaev}, K.~{Byrum}, A.~{Cannon}, O.~{Celik},
  A.~{Cesarini}, Y.~C. {Chow}, L.~{Ciupik}, P.~{Cogan}, P.~{Colin}, W.~{Cui},
  R.~{Dickherber}, C.~{Duke}, S.~J. {Fegan}, J.~P. {Finley}, G.~{Finnegan},
  P.~{Fortin}, L.~{Fortson}, A.~{Furniss}, N.~{Galante}, D.~{Gall}, K.~{Gibbs},
  G.~H. {Gillanders}, S.~{Godambe}, J.~{Grube}, R.~{Guenette}, G.~{Gyuk},
  D.~{Hanna}, J.~{Holder}, D.~{Horan}, C.~M. {Hui}, T.~B. {Humensky},
  A.~{Imran}, P.~{Kaaret}, N.~{Karlsson}, M.~{Kertzman}, D.~{Kieda},
  J.~{Kildea}, A.~{Konopelko}, H.~{Krawczynski}, F.~{Krennrich}, M.~J. {Lang},
  S.~{Lebohec}, G.~{Maier}, S.~{McArthur}, A.~{McCann}, M.~{McCutcheon},
  J.~{Millis}, P.~{Moriarty}, R.~{Mukherjee}, T.~{Nagai}, R.~A. {Ong}, A.~N.
  {Otte}, D.~{Pandel}, J.~S. {Perkins}, F.~{Pizlo}, M.~{Pohl}, J.~{Quinn},
  K.~{Ragan}, L.~C. {Reyes}, P.~T. {Reynolds}, E.~{Roache}, H.~J. {Rose},
  M.~{Schroedter}, G.~H. {Sembroski}, A.~W. {Smith}, D.~{Steele}, S.~P.
  {Swordy}, M.~{Theiling}, S.~{Thibadeau}, A.~{Varlotta}, V.~V. {Vassiliev},
  S.~{Vincent}, R.~G. {Wagner}, S.~P. {Wakely}, J.~E. {Ward}, T.~C. {Weekes},
  A.~{Weinstein}, T.~{Weisgarber}, D.~A. {Williams}, S.~{Wissel}, M.~{Wood},
  and B.~{Zitzer}.
\newblock {A connection between star formation activity and cosmic rays in the
  starburst galaxy M82}.
\newblock {\em \nat}, 462(7274):770--772, December 2009.

\bibitem{HESS:2018yqa}
H.~Abdalla et~al.
\newblock {The starburst galaxy NGC 253 revisited by H.E.S.S. and Fermi-LAT}.
\newblock {\em Astron. Astrophys.}, 617:A73, 2018.

\bibitem{MAGIC:2019fvw}
V.~A. Acciari et~al.
\newblock {Constraints on gamma-ray and neutrino emission from NGC 1068 with
  the MAGIC telescopes}.
\newblock {\em Astrophys. J.}, 883:135, 2019.

\end{thebibliography}

\end{document}